\documentclass[11pt]{article}
\usepackage{amsmath}
\usepackage{amssymb}
\usepackage{graphicx}
\usepackage{enumerate}
\usepackage{authblk}
\usepackage{cite}
\usepackage{color}
\usepackage[margin=1.0in]{geometry}
\normalsize \baselineskip 16pt
\newcommand {\beq} {\begin{equation}}
\newcommand {\eeq} {\end{equation}}

\begin{document}
\title{Solitary waves and kinks in FPU lattices with soft-hard-soft trilinear interactions}
\author[1]{Anna Vainchtein}
\author[2]{Lev Truskinovsky}
\affil[1]{\small Department of Mathematics, University of Pittsburgh, Pittsburgh, Pennsylvania 15260, USA, \texttt{aav4@pitt.edu}}
\affil[2]{\small PMMH, CNRS--UMR 7636, ESPCI ParisTech, 10 Rue Vauquelin, Paris, 75005, France \texttt{lev.truskinovsky@espci.fr}}

\maketitle

\begin{abstract}

We consider a version of the classical Hamiltonian Fermi-Pasta-Ulam (FPU) problem with a trilinear force-strain relation of soft-hard-soft type that is in general non-symmetric. In addition to the classical spatially localized solitary waves, such hardening-softening model also exhibits supersonic kinks and finite-amplitude, spatially delocalized flat-top solitary waves that acquire the structure of a kink-antikink bundle when their velocity approaches the kink limit. Exploiting the fact that traveling waves are periodic modulo shift by a lattice spacing, we compute these solutions as fixed points of the corresponding nonlinear map and investigate how their properties depend on the parameter measuring the asymmetry of the problem. In a particularly interesting case when one of the soft regimes has zero elastic modulus, we obtain explicit solutions for sufficiently slow solitary waves. In contrast to conventional delocalization in the sonic limit, these compact structures mounted on a constant background become localized at the lattice scale as their velocity tends to zero. Numerical simulations of Riemann-type initial value problem in this degenerate model show the emergence of Whitham shocks that involve periodic trains of solitary waves. We investigate stability of the obtained solutions using direct numerical simulations and Floquet analysis. We also obtain explicit solutions for a quasicontinuum model that captures some important features of the discrete problem.

\end{abstract}

\section{Introduction}
Pulse-shaped solitary waves constitute an important class of traveling waves in nonlinear systems. These localized dynamic coherent structures emerge in discrete and continuum mechanical systems due to the interplay of dispersion and nonlinearity.  Stable solitary waves  play an important role as building blocks in developing dynamical patterns in various nonlinear mechanical systems, ranging from granular crystals to metamaterials, that are used in applications exploiting structural nonlinearities at the scale of the  periodicity cell \cite{yasuda2020transition,raney2016stable,kochmann2017exploiting,zhang2019programmable}. Artificially created materials of this type   can now manipulate localized mechanical signals, and the ensuing control of solitary waves is used for mechanical energy transmission,  encryption of mechanical information  and even activation of mechanical robots \cite{bertoldi2017flexible,yasuda2019origami}.

Solitary waves in discrete mechanical systems and their continuum KdV-type approximations were first discovered in the pioneering work \cite{ZK65} that explained the seemingly paradoxical results of the numerical investigations in \cite{fermi1955studies} of the nonintegrable Hamiltonian Fermi-Pasta-Ulam (FPU) lattice, a mass-spring chain with nonlinear nearest-neighbor interactions \cite{berman2005fermi,gallavotti2007fermi}. In subsequent studies solitary waves emerged as localized, non-topological and  non-dissipative coherent structures that move with supersonic speeds and form continuous families  \cite{remoissenet2013waves, newell1985solitons, fokas2012important,Vainchtein22,ablowitz2011nonlinear}.  The most well studied case of solitary waves in discrete FPU system is when the springs are characterized  by  force-strain relation  of either hardening or softening type, as for instance in the case of  $\alpha$-FPU  system with quadratic nonlinearity. While weak solitary waves  in such systems can be characterized as  low-amplitude, completely delocalized and almost linear waves, very strong solitary waves emerge as  maximally localized,  lattice scale  anti-continuum mechanical signals.  The  most analytically transparent setting in this class of problems is the \emph{bilinear}, soft-hard  model introduced already in the original FPU study \cite{fermi1955studies}.  In the present paper we consider a \emph{trilinear}, soft-hard-soft, generalization of this classical model.  Using exact solutions available in this case we show that, even without compromising the convexity of the energy, the resulting hardening-softening system exhibits non-classical physical effects.

Specifically, we consider the prototypical discrete FPU chain whose mechanical response is represented by three linear elastic regimes which we characterize as soft, hard and again soft. This implies that the conventional hardening soft-hard response is eventually taken over by a softening hard-soft regime. No symmetry is assumed regarding the two soft regimes which in high contrast limit would be  characterized by drastically different elastic moduli. Our goal is to take advantage of the fact that the addition of the second soft regime leads to the emergence in such FPU system of rather peculiar delocalized finite-amplitude flat-top solitary wave solutions, which
are intimately connected to the nontopological supersonic kink solutions. Due to the piecewise linear nature of the problem, both kinks and solitary waves can be studied analytically in a quasicontinuum approximation of the discrete system.

The choice of soft-hard-soft interactions is inspired by stress-strain laws in a range of soft biological tissues  from  skin  to muscles \cite{Yasenchuk21}.  For instance, in tendons and  ligaments the hardening stage of the  mechanical  response can  be linked to the  straightening of crimped collagen fibers while the softening  stage may be due to the beginning of the distributed microscopic fracturing of these fibers \cite{Yasenchuk21,Sensini18}.
Hardening to softening transition is also ubiquitous in elastomeric molecular composites \cite{Millereau18} and is sometimes mimicked in NiTi mesh implants \cite{Yasenchuk21}.

The question of existence of traveling waves in a hardening-softening FPU system has been already addressed in the literature  \cite{Iooss00,HerrmannRademacher10,Herrmann11,Gorbushin19}. Two recent papers discussed the relation between  solitary waves  and nontopological kinks in such systems. In one of them \cite{Gorbushin21} the force-elongation relation was taken in a bilinear,  soft-soft form  with a  degenerate infinitely hard response in between.  In the other \cite{VT24}  the mechanical response was chosen to be cubic with symmetric softening and hardening regimes. Both models produced a coherent description of the families of solitary waves that in a special velocity limit feature formation of supersonic kinks, or superkinks \cite{VT24, Gorbushin22}. As this limit is approached, the waves increase in width and acquire a flat-top finite-amplitude structure of a kink-antikink bundle.

However, the emerging picture remains incomplete. Thus, the fact that the bilinear model replaced the hard section of the constitutive response by an infinitely hard one did not leave any space for the internal degrees of freedom governing the energy transfer inside the core regions of both solitary waves and superkinks. In particular, this resulted in an unrealistic prediction that such solutions may propagate with arbitrarily large speeds. In addition, to enable a simple solution procedure based on the Fourier transform, the two soft regions in the bilinear model considered in \cite{Gorbushin21} were taken to be fully symmetric. For the same reason of analytical simplicity the cubic model studied in \cite{VT24} was also chosen to be overly symmetric, which made the repertoire  of possible physical effects somewhat limited  while concealing some interesting special cases such as the high-contrast case when one of the two soft regimes has zero sound speed (a ``sonic vacuum" \cite{Vainchtein22}).  Of course, the  important advantage of the constitutive  choices made in \cite{Gorbushin21, VT24} was that in the cubic case a quasicontinuum (QC) approximation of the discrete FPU problem yields explicit solutions, while in the bilinear case analytical solutions can be found for both discrete and QC problems.

To complement the existing studies, we present in this paper the still missing discussion of the non-symmetric trilinear case. First, it allows us to study the case of radically different soft regimes including the limit when one of them becomes elastically degenerate. Such limit, which resembles granular response, is of interest as the model becomes non-linearizable in the corresponding strain range, and nonlinearity becomes essential. In particular, due to such degeneracy the linear waves disappear, while the conventional dispersive shock waves are replaced by Whitham shocks involving trains of compact solitary waves. The second advantage of the trilinear model is the possibility of opening up the core region of both solitary waves and kinks which is then described by a separate degree of freedom that can have its own evolution. This allows one to study interaction between the different parts of the traveling wave and results in a realistic finite velocity limit for its propagation.

We start by considering a QC approximation\cite{collins1981quasicontinuum,Rosenau86,kevrekidis2002continuum,feng2004quasi} of the discrete FPU model for which all of these effects can be demonstrated using explicit traveling wave solutions in the form of kinks and solitary waves. In particular, we show that asymmetry of the problem, measured by the non-unit ratio of the elastic moduli in the soft regimes, has a significant effect on the velocity ranges of the existence of kink solutions, as well as the values and asymptotic behavior of their limiting strains as functions of the velocity. This, in turn, affects the limiting amplitudes and velocity ranges of the associated solitary waves. We also discuss the asymptotic behavior of the obtained solutions near the boundaries of the velocity interval. In particular, we show that in the case when one of the soft regimes has zero modulus, the compressive solitary waves have a nontrivial sonic limit.

We then follow the approach in \cite{Aubry09,Vainchtein20,James21,VT24} and take advantage of the fact that traveling wave solutions of the discrete problem are periodic modulo shift by one lattice spacing to obtain such solutions of the discrete problem as fixed points of a nonlinear map. While this procedure generally requires numerical iterations, in the important case when one of the soft regimes becomes degenerate, sufficiently slow discrete solitary waves can be also computed analytically. This allows us to corroborate our numerical procedure and reveal the compact (modulo constant background) nature of the solutions in this limiting case, which is not captured by the QC model. Further comparison with traveling wave solutions of the discrete problem shows that while the QC model captures them qualitatively, the quantitative agreement is fairly good for superkinks but exists primarily near sonic and kink velocity limits for solitary waves. The discrepancy between the solutions away from these limits depends on the nature of the wave (compressive or tensile) and the value of the asymmetry parameter.

Floquet analysis of the linear stability of obtained solitary waves in the non-degenerate cases, which is enabled by their periodicity-modulo-shift \cite{Vainchtein20,Cuevas17,Xu18} and takes advantage of the piecewise linear nature of problem, shows that near-sonic solutions with velocity below a certain threshold are unstable. Numerical simulations initiated by the unstable waves perturbed along the corresponding eigenmode show that the instability leads in the system approaching an apparently stable wave above the threshold. Effective stability of such waves above the threshold is confirmed by direct numerical simulations, which also show robust propagation of superkinks and solitary waves in the degenerate case at the prescribed velocities. This is further corroborated by the Riemann-type simulations with piecewise-constant initial data, which also reveal other interesting phenomena, such as formation of Whitham shocks involving periodic trains of solitary waves in the degenerate case.

The paper is organized as follows. In Sec.~\ref{sec:formulation} we formulate the problem and discuss the general properties of the two types of traveling waves. Explicit solutions for the QC model are constructed and discussed in Sec.~\ref{sec:QC}. In Sec.~\ref{sec:discrete} we describe the procedure for computing traveling wave solutions of the discrete problem, derive explicit solutions for sufficiently slow compressive waves in the degenerate case, and compare the obtained solutions with their QC counterparts. Stability of kinks and solitary waves is discussed in Sec.~\ref{sec:stab_super} and Sec.~\ref{sec:stab_SW}, respectively. Concluding remarks can be found in Sec.~\ref{sec:conclusions}. Some technical results are contained in Appendix~\ref{app:coefs}.

\section{Problem formulation}
\label{sec:formulation}
\begin{figure}
\centering
\includegraphics[width=\textwidth]{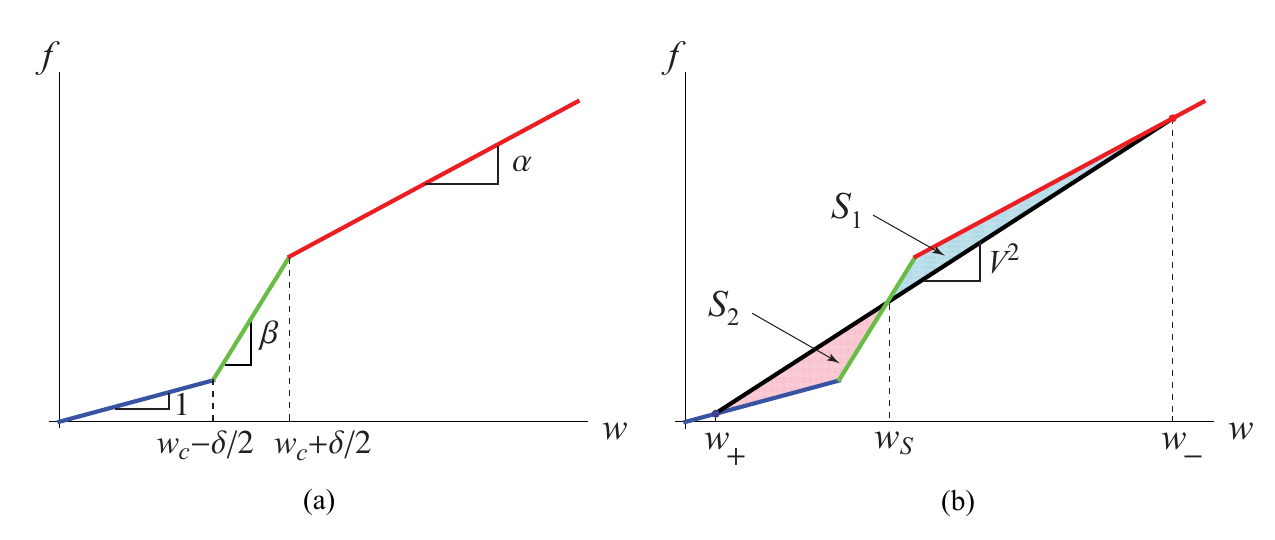}
\caption{\footnotesize (a) Trilinear soft-hard-soft interaction force $f(w)$ with slopes $1$, $\beta$ and $\alpha$ along the blue (soft), green (hard) and red (soft) segments, respectively; (b) Rayleigh line connecting $(w_{+},f(w_{+}))$ and $(w_{-},f(w_{-}))$ (black) for a superkink transition wave with limiting states $w_{\pm}$ and supersonic velocity $V$ satisfying $\max\{1,\alpha\}<V^2<\beta$. The two shaded areas cut by the Rayleigh line at $w=w_S$ are equal: $S_1=S_2$ (see the text for details).}
\label{fig:superkink}
\end{figure}
Consider a one-dimensional chain of identical masses interacting with their nearest neighbors. The dimensionless governing equations are
\beq
\ddot{u}_n=f(u_{n+1}-u_n)-f(u_n-u_{n-1}),
\label{eq:FPU}
\eeq
where $u_n(t)$ is the displacement of $n$th particle at time $t$, $\ddot{u}_n(t)=u_n''(t)$, and $f(w)$ is the nonlinear interaction force associated with the interaction potential $\Phi(w)=\int_0^s f(s)ds$.
Introducing the strain variables $w_n=u_n-u_{n-1}$, we can rewrite \eqref{eq:FPU}
in the form
\beq
\ddot{w}_n=f(w_{n+1})-2f(w_n)+f(w_{n-1}).
\label{eq:FPUstrain}
\eeq
Our main assumption concerns the choice of the particle interactions in the trilinear soft-hard-soft form:
\beq
f(w)=\begin{cases} w, & w \leq w_c-\frac{\delta}{2}\\
                  w_c-\frac{\delta}{2}+\beta(w-w_c+\frac{\delta}{2}), & |w-w_c| \leq \frac{\delta}{2}\\
                  \begin{cases} \alpha(w-b), & \alpha >0\\
                  w_c-\frac{\delta}{2}+\beta\delta, & \alpha=0
                  \end{cases}, & w \geq w_c+\frac{\delta}{2}
     \end{cases},
\label{eq:trilinear}
\eeq
where we assume
\beq
0 \leq \alpha < \beta, \quad \beta>1, \quad \delta>0, \quad w_c-\dfrac{\delta}{2}>0
\label{eq:param}
\eeq
and define, for $\alpha>0$,
\beq
b=w_c+\dfrac{\delta}{2}-\dfrac{1}{\alpha}\left(w_c-\dfrac{\delta}{2}+\beta\delta\right).
\label{eq:b}
\eeq
As illustrated in Fig.~\ref{fig:superkink}(a), $f(w)$ is a continuous piecewise linear function consisting of three linear segments with slopes $1$ (soft, blue segment), $\beta$ (hard, green segment) and $\alpha$ (soft, red segment). The width of the intermediate (hard) green segment is controlled by the parameter $\delta>0$. In the limit $\delta \to 0$ its slope $\beta$ tends to infinity, and we obtain a bilinear function with a jump discontinuity at $w=w_c$ that was considered in \cite{Gorbushin22} for $\alpha>1$ and in \cite{Gorbushin19,Gorbushin21} for $\alpha=1$.

Note that since $\beta>1$ and $\alpha<\beta$, $f(w)$ has a \emph{hardening-softening} form that changes from convex to concave at any point along the green segment in Fig.~\ref{fig:superkink}(a). In the special case $\alpha=0$ the interaction force saturates to a constant value above $w_c+\delta/2$. This inelastic state corresponds to zero sound speed (``sonic vacuum'').

In this paper we are interested in traveling waves that connect stable equilibrium states of the system \eqref{eq:FPUstrain}, with constant strains $w_\pm$ such that $f'(w_{\pm}) \geq 0$, and propagate with velocity $V$ that is supersonic with respect to both limiting states:
\beq
w_n(t)=w(\xi), \quad \xi=n-Vt,
\label{eq:TWansatz}
\eeq
where
\beq
\lim_{\xi \to \pm \infty} w(\xi) = w_{\pm}
\label{eq:BCs}
\eeq
and $V^2>f'(w_{\pm})$.
The function $w(\xi)$ must thus satisfy the advance-delay differential equation
\beq
V^2w''(\xi)=f(w(\xi+1))-2f(w(\xi))+f(w(\xi-1)).
\label{eq:TW_discrete}
\eeq

\subsection{Superkinks}
Suppose the traveling wave (TW) is a monotone front connecting two \emph{different} limiting states, $w_{-} \neq w_{+}$ in the blue and red segments, as shown in Fig.~\ref{fig:superkink}. Such transition waves have been classified in \cite{Gorbushin22} as supersonic kinks, or \emph{superkinks}.

One can show \cite{Serre07,Aubry09,HerrmannRademacher10,Herrmann11,Gorbushin19,Gorbushin21,Gorbushin22} that in addition to the classical Rankine-Hugoniot jump condition
\beq
f(w_{+})-f(w_{-})=V^2(w_{+}-w_{-}),
\label{eq:Rline}
\eeq
which states that the slope of the \emph{Rayleigh line} connecting $(w_+,f(w_+))$ and $(w_{-},f(w_{-}))$ equals $V^2$, as shown in Fig.~\ref{fig:superkink}, such solutions must satisfy the condition
\beq
\Phi(w_{+})-\Phi(w_{-})-\dfrac{1}{2}(w_{+}-w_{-})(f(w_{+})+f(w_{-}))=0.
\label{eq:zero_G}
\eeq
This additional condition constitutes a \emph{kinetic relation} for a superkink. More precisely, it states that the \emph{driving force} $G=\Phi(w_{+})-\Phi(w_{-})-\frac{1}{2}(w_{+}-w_{-})(f(w_{+})+f(w_{-}))$ \cite{Trusk87} on the moving front is zero, and thus
there is no dissipation associated with its motion. Geometrically it means that the two areas cut by the Rayleigh line from $f(w)$ must be equal, as shown in Fig.~\ref{fig:superkink}(b). Due to the trilinear form \eqref{eq:trilinear} of $f(w)$, it follows that the superkink velocity $V$ must satisfy
\beq
\max\{1,\alpha\}<V^2<\beta.
\label{eq:V_limits}
\eeq

The conditions \eqref{eq:Rline} and \eqref{eq:zero_G} imply that in the case of superkinks, only one of the values $w_{-}$, $w_+$ and $V$ can be prescribed independently. In particular, they determine $|V|$ and $w_{-}$ for a given $w_+$. Global existence of superkinks in the FPU problem with smooth hardening-softening interactions was proved in \cite{HerrmannRademacher10,Herrmann11} under the area condition \eqref{eq:zero_G}. Local analysis in \cite{Iooss00} has shown that for smooth $f(w)$ small-amplitude superkinks bifurcate from local maxima of $f'(w)$ connecting convex and concave parts of $f(w)$. Exact superkink solutions in the problem with bilinear interactions ($\delta=0$ in \eqref{eq:trilinear}) were constructed in \cite{Gorbushin19,Gorbushin21,Gorbushin22}. Note that for each superkink solution propagating with velocity $V$, there exists a solution of the same form but velocity $-V$. In addition, for each kink solution with $w_{-}>w_{+}$, i.e., a front with $w'(\xi)<0$, there is an \emph{antikink} solution $\tilde{w}(\xi)=w(-\xi)$ with the limiting states interchanged, so that $\tilde{w}'(\xi)>0$, and the same velocity. Thus, it suffices to consider kink solutions with $V>0$.

\subsection{Solitary waves}
As discussed in \cite{Gorbushin19,Gorbushin21,VT24}, superkinks are closely related to \emph{solitary waves}, pulse-like solutions of \eqref{eq:TW_discrete} connecting \emph{identical} limiting states, $w_{-}=w_{+}$, and propagating with supersonic velocities. Existence of solitary wave (SW) solutions has been shown in \cite{Friesecke94}; see also \cite{Vainchtein22} for a recent review. Note that such solutions automatically satisfy \eqref{eq:Rline} and \eqref{eq:zero_G}. Solitary waves can be \emph{tensile}, $w(\xi)>w_+$, or \emph{compressive}, $w(\xi)<w_+$. Similar to the superkinks, for each solitary wave moving with velocity $V$, there is a wave of the same form moving with velocity $-V$, so it suffices to consider positive velocities.

Importantly, the speed of the solitary wave solutions that tend to $w_+$ at plus and minus infinity is bounded from below by the sonic limit and from above by the superkink speed: $f'(w_+)<V^2<V_{SK}^2$, where $V_{SK}$ is the velocity of the superkink with the state $w_+$ ahead. As the superkink limit is approached, solitary wave solutions increase in amplitude and become wider and more flat in the middle, with the two boundary layers on the left and on the right that approximate monotone superkink solutions. Thus, for velocities just below the superkink limit, solitary waves acquire a structure where a kink and an antikink move in tandem. This will be further illustrated by explicit solutions constructed in the next section for a QC model. Solitary waves of this type, sometimes referred to as ``flat-top solitons", have been recently obtained for first-order nonlinear systems including the extended Gardner-like equations \cite{RosenauOron20,RosenauOron22} and oscillator chains \cite{RosenauPikovsky20,RosenauPikovsky21}. In the context of the FPU problem, such solutions and the limiting superkinks were obtained in \cite{VT24} for cubic interactions and in \cite{Gorbushin19,Gorbushin21} for the special case of bilinear interactions with equal slopes that enables analytical treatment of the discrete problem.

As shown in \cite{Gorbushin19,Gorbushin21,VT24}, one can also construct solitary wave solutions above the superkink limit. Such solutions have velocity-dependent background state at infinity and tend to a bound kink-antikink structure as the superkink limit is approached from above. In this work, however, we limit our attention to solitary waves below the superkink limit.

\section{Exact solutions for a quasicontinuum model}
\label{sec:QC}
In view of the complexity of the original discrete problem represented by an infinite system \eqref{eq:FPU} of nonlinear ordinary differential equations, we first turn first to a model representing its analytically transparent QC approximation that yields exact TW solutions. The QC model we consider is described by
the regularized Boussinesq partial differential equation
\beq
u_{tt}-\dfrac{1}{12}u_{xxtt}=(f(u_x))_x,
\label{eq:Bous}
\eeq
which can be obtained from \eqref{eq:FPU} using the $(2,2)$ Pad\'e approximation, $4\sin^2(k/2) \approx k^2/(1+k^2/12)$, of the discrete Laplacian in Fourier space \cite{Rosenau86}. The associated
Lagrangian density
\beq
\mathcal{L}=\dfrac{1}{2}\left(u_t^2+\dfrac{1}{12}u_{tx}^2\right)-\Phi(u_x),
\label{eq:Lagrangian}
\eeq
contains an additional ``microkinetic'' energy term $(1/24)u_{tx}^2$.

One can show \cite{VT24} that in this model the traveling wave equation for $w(\xi)=u_x(x,t)$, $\xi=x-Vt$, reduces to
\beq
-\dfrac{V^2}{12}w''+V^2 w-f(w)=V^2 w_{+}-f(w_+),
\label{eq:TW_QC_gen2}
\eeq
where the boundary condition at $\xi \to \infty$ in \eqref{eq:BCs} was used.
Together with the boundary condition at $\xi \to -\infty$ in \eqref{eq:BCs}, this yields the Rankine-Hugoniot condition \eqref{eq:Rline}. Integrating \eqref{eq:TW_QC_gen2} results in the first-order ordinary differential equation
\beq
-\dfrac{V^2}{24}(w')^2=\Phi(w)-\Phi(w_{+})-f(w_{+})(w-w_{+})-\dfrac{V^2}{2}(w-w_{+})^2,
\label{eq:TW_QC_gen3}
\eeq
where the boundary condition at $\xi \to \infty$ in \eqref{eq:BCs} was used again.
In view of the boundary condition at $\xi \to -\infty$ in \eqref{eq:BCs}, this yields
\[
\Phi(w_{-})-\Phi(w_{+})-f(w_{+})(w_{-}-w_{+})-\dfrac{V^2}{2}(w_{-}-w_{+})^2=0,
\]
which together with \eqref{eq:Rline} implies that the equal-area condition \eqref{eq:zero_G}. For a superkink solution of \eqref{eq:TW_QC_gen2}, the limiting states $w_{\pm}$ satisfy \eqref{eq:Rline} and \eqref{eq:zero_G} for a given $V$. For solitary waves, we can independently prescribe the background state $w_+$ and supersonic velocity $V$ with magnitude below the superkink limit.

In the case of trilinear interactions \eqref{eq:trilinear}, equation \eqref{eq:TW_QC_gen2} can be solved analytically in each segment (blue, green and red) where the elastic modulus is constant and the corresponding ordinary differential equation is linear. The obtained solutions can then be matched using the continuity conditions, as described below.

\subsection{Superkinks}
Consider first the superkink solutions \eqref{eq:TW_QC_gen2} with $f(w)$ given by \eqref{eq:trilinear} that connect the states $w_+$ and $w_{-}$ in the blue and red segments in Fig.~\ref{fig:superkink}, respectively ($w_+<w_c-\delta/2$ and $w_{-}>w_c+\delta/2$) and propagate with velocity $V>0$ that satisfies \eqref{eq:V_limits}. Observe that in this case we have $f(w_+)=w_+$, so \eqref{eq:TW_QC_gen2} simplifies to
\beq
-\dfrac{V^2}{12}w''+V^2 w-f(w)=(V^2-1)w_{+},
\label{eq:TW_QC}
\eeq
while \eqref{eq:Rline} becomes
\beq
w_{-}=\dfrac{(V^2-1)w_{+}-\alpha b}{V^2-\alpha}
\label{eq:RH}
\eeq
for $\alpha>0$ and
\beq
w_{-}=\dfrac{(V^2-1)w_{+}+w_c+\beta\delta-\delta/2}{V^2}
\label{eq:RH_alpha0}
\eeq
for $\alpha=0$.
We seek monotone kink solutions such that $w(\xi)>w_c+\delta/2$ (red linear segment of $f(w)$ in Fig.~\ref{fig:superkink}) for $\xi<-z$, where $z>0$ is to be determined, $|w(\xi)-w_c|<\delta/2$ (green segment) for $-z<\xi<z$ and $w(\xi)<w_c-\delta/2$ (blue segment) for $\xi>z$. Solving the corresponding linear equations in each interval yields
\beq
w(\xi)=\begin{cases} w_{+}+Ae^{-r\xi}, & \xi \geq z,\\
                     w_S+B\cos(q\xi)+C\sin(q\xi), & |\xi| \leq z,\\
                     w_{-}+De^{s\xi}, & \xi \leq -z,
       \end{cases}
\label{eq:soln_form}
\eeq
where $w_{-}$ is related to $w_{+}$ via \eqref{eq:RH} for $\alpha>0$ and \eqref{eq:RH_alpha0} for $\alpha=0$,
\beq
w_S=\dfrac{(w_c-\delta/2)(\beta-1)-w_+(V^2-1)}{\beta-V^2}
\label{eq:wS}
\eeq
is the intersection of $f(w)$ and the Rayleigh line in the hard (green) linear regime (see Fig.~\ref{fig:superkink}(b)),
and the roots $r$, $q$ and $s$ are given by
\beq
r=\dfrac{\sqrt{12(V^2-1)}}{V}, \quad q=\dfrac{\sqrt{12(\beta-V^2)}}{V}, \quad s=\dfrac{\sqrt{12(V^2-\alpha)}}{V}.
\label{eq:roots}
\eeq
Note that at $\alpha=0$ we have $s=\sqrt{12}$.

Before providing further details about the solution \eqref{eq:soln_form}, we discuss a physical interpretation of its structure.
One can think of the frontal part of the superkink solution ($\xi>a$, $|a|<z$) as a portion of the structured shock wave propagating with velocity $V$ that has the strain $w_+$ in front and oscillations around the average strain $w_S$ with wave number $q$. In the non-dispersive continuum limit the transition layer and the oscillations disappear from the shock's structure, and it becomes a moving discontinuity that dissipates energy at the rate $V S_2$, where $S_2$ is the shaded pink area in Fig.~\ref{fig:superkink}(b). As discussed in \cite{Gorbushin22}, this energy release rate can equivalently be computed on the microscopic level by accounting for the energy radiated in form of the dispersive wave propagating behind the shock. Note, however, that only a portion of this shock solution is included in \eqref{eq:soln_form}. Meanwhile, the back part of the superkink ($\xi<a$) can be represented by a portion of the structured shock wave that has oscillations around $w_S$ ahead and $w_{-}$ behind. This second wave is not an admissible shock wave because it is supersonic with respect to the state behind and thus violates the Lax condition. It absorbs energy at the rate $V S_1$, where $S_1$ is the blue shaded area in Fig.~\ref{fig:superkink}(b), which, as we discussed above, equals $S_2$, the rate at which the energy is released. A superkink can thus be thought of as a bundle of admissible and inadmissible shock waves, where the energy released in the front is transported to the back, where it is absorbed, by the mode $q$. Indeed, observing that the dispersion relation in the hard linear regime is given by $\omega^2=\beta k^2/(1+k^2/12)$ in the QC model, one can show that the energy is carried with the group velocity $\omega'(q)=V^3/\beta$, which is less than the phase velocity $V$ since $V^2<\beta$, and thus the energy is transported from front to back. Inside the superkink bundle, the energy sink (inadmissible shock wave) is stabilized by the elastic radiation from the energy source (admissible shock wave), which is sometimes called the ``feeding wave'' \cite{slepyan2001feeding,gorbushin2020frictionless}.

\begin{figure}
\centering
\includegraphics[width=\textwidth]{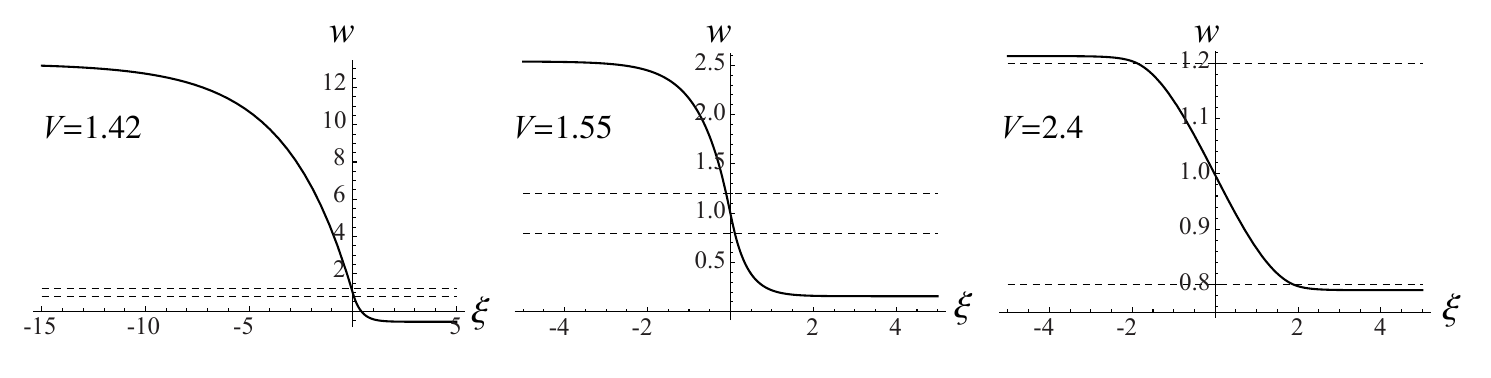}
\caption{\footnotesize Strain profiles at different velocities when $\alpha=2$, $\delta=0.4$, $b=-0.4$, $w_c=1$. The left panel shows the superkink traveling with velocity just above the lower limit $\sqrt{\alpha} \approx 1.41$, and the right panel the one with velocity slightly below the upper limit $\sqrt{\beta} \approx 2.45$. Dashed horizontal lines mark $w=w_c \pm \delta/2$.}
\label{fig:strains}
\end{figure}

To find the six unknown variables in $z$, $w_{+}$ and the coefficients $A$, $B$, $C$ and $D$ \eqref{eq:soln_form}, for given $V$, we apply the continuity conditions for $w(\xi)$ and $w'(\xi)$ at $\xi=\pm z$ (four conditions) and the two switch conditions $w(\pm z)=w_c \pm \delta/2$. In the generic case  $0 \leq \alpha \neq 1$ this yields
\beq
w(\xi)=w_{+}+\dfrac{\delta e^{-r(\xi-z)}}{\alpha-1}\left[\beta-\alpha-\dfrac{\sqrt{(\beta-1)(\beta-\alpha)(V^2-\alpha)}}{\sqrt{V^2-1}}\right],
\quad \xi \geq z,
\label{eq:soln1}
\eeq
\beq
\begin{split}
w(\xi)&=\dfrac{\delta\sqrt{(\beta-\alpha)(\beta-1)}}{\sqrt{(\beta-\alpha)(V^2-1)}+\sqrt{(\beta-1)(V^2-\alpha)}}\cos(q\xi-\phi)\\
&+\dfrac{(\beta-1)(w_c-\delta/2)-(V^2-1)w_{+}}{\beta-V^2}, \quad |\xi| \leq z,
\end{split}
\label{eq:soln2}
\eeq
where
\[
\phi=\pi\theta(\alpha-1)+\arctan\dfrac{\sqrt{(\beta-\alpha)(V^2-1)}+\sqrt{(\beta-1)(V^2-\alpha)}}{\sqrt{\beta-V^2}(\sqrt{\beta-1}-\sqrt{\beta-\alpha})},
\]
and
\beq
w(\xi)=w_{-}+\dfrac{\delta e^{s(\xi+z)}}{\alpha-1}\left[\beta-1-\dfrac{\sqrt{(\beta-1)(\beta-\alpha)(V^2-1)}}{\sqrt{V^2-\alpha}}\right],
\quad \xi \leq -z.
\label{eq:soln3}
\eeq
Here
\beq
\begin{split}
z&=\dfrac{V}{4\sqrt{3}\sqrt{\beta-V^2}}\bigg\{\arctan\dfrac{\sqrt{\beta-V^2}(\sqrt{V^2-\alpha}+\sqrt{V^2-1})}{\beta-V^2-\sqrt{(V^2-\alpha)(V^2-1)}}\\
&+\pi\theta(\sqrt{(V^2-\alpha)(V^2-1)}-\beta+V^2)\bigg\},
\end{split}
\label{eq:z}
\eeq
where $\theta(x)=1$ for $x>0$ and zero otherwise, and the limiting states are given by
\beq
w_{+}=w_c+\dfrac{\delta}{2(\alpha-1)}\left\{1+\alpha-2\beta+\dfrac{2\sqrt{(\beta-\alpha)(\beta-1)(V^2-\alpha)}}{\sqrt{V^2-1}}\right\}
\label{eq:wplus}
\eeq
and
\beq
w_{-}=w_c+\dfrac{\delta}{2(\alpha-1)}\left\{1+\alpha-2\beta+\dfrac{2\sqrt{(\beta-\alpha)(\beta-1)(V^2-1)}}{\sqrt{V^2-\alpha}}\right\}.
\label{eq:wminus}
\eeq
Some examples of strain profiles are shown in Fig.~\ref{fig:strains}.
The particle velocity is given by $v(\xi)=-V w(\xi)$.

In the symmetric case $\alpha=1$ (equal slopes of the red and blue segments), we have $s=r$, and the solution is given by
\beq
w(\xi)=\begin{cases} w_{+}+\dfrac{\delta (\beta-V^2)}{2(V^2-1)}e^{-r(\xi-z)}, & \xi \geq z\\
                     \dfrac{(\beta-1)(w_c-\delta/2)-(V^2-1)w_{+}}{\beta-V^2}-\dfrac{\delta\sqrt{\beta-1}}{2\sqrt{V^2-1}}\sin(q\xi), & |\xi| \leq z\\
                     w_{-}-\dfrac{\delta (\beta-V^2)}{2(V^2-1)}e^{r(\xi+z)}, & \xi \leq -z,
        \end{cases}
\label{eq:soln_alpha1}
\eeq
with
\beq
z=\dfrac{V}{4\sqrt{3}\sqrt{\beta-V^2}}\bigg\{\arctan\dfrac{2\sqrt{\beta-V^2}\sqrt{V^2-1}}{\beta-2V^2+1}+\pi\theta(2V^2-\beta-1)\biggr\}
\label{eq:z_alpha1}
\eeq
and
\beq
w_{\pm}=w_c \mp \dfrac{\delta(\beta-1)}{2(V^2-1)}=w_c \pm \dfrac{b}{2(V^2-1)},
\label{eq:wpm_alpha1}
\eeq
so that the limiting strains are independent of $\delta$ (note that \eqref{eq:b} implies that $b=\delta(1-\beta)<0$ in this case).

\begin{figure}[h]
\centering
\includegraphics[width=\textwidth]{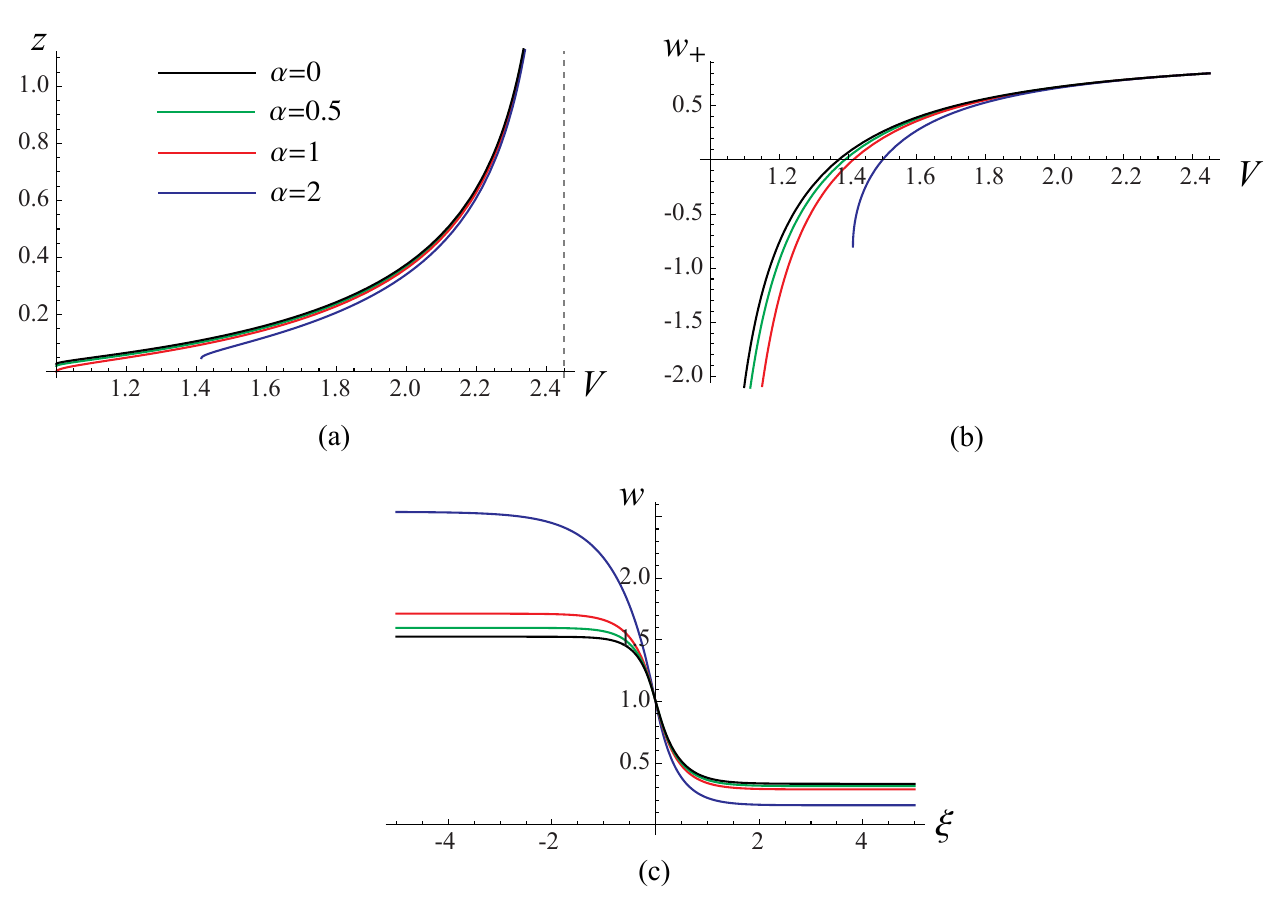}
\caption{\footnotesize (a) The functions $z(V)$, which tend to infinity as $V \to \sqrt{\beta}$ (dashed vertical line), for different $\alpha$. For $\alpha>1$ (blue curve), the lower velocity bound is $\sqrt{\alpha}$. (b) Plots of $w_{+}(V)$. Note that $w_+$ tends to $-\infty$ as $V \to 1$ for $\alpha \leq 1$ (black, green and red curves), while for $\alpha>1$ (blue curve) it has a finite value at $V=\sqrt{\alpha}$. (c) Strain profiles at $V=1.55$. The legend in panel (a) applies to all four panels. Here  $\beta=6$, $\delta=0.4$ and $w_c=1$.}
\label{fig:vary_alpha}
\end{figure}

The effect of $\alpha$ on $z(V)$, $w_{+}(V)$ and the solution profiles is shown in Fig.~\ref{fig:vary_alpha}.
Recall that by \eqref{eq:V_limits} the upper velocity limit is $\sqrt{\beta}$ in all cases. As this limit is approached, $z$ tends to infinity:
\beq
z \approx \dfrac{\pi\sqrt{\beta}}{8\sqrt{3}\sqrt{\beta-V^2}} \quad \text{as $V \to \sqrt{\beta}$},
\label{eq:z_upper_limit}
\eeq
while the two limiting strains approach the boundaries of the intermediate linear segment: $w_{\pm} \to w_c \mp \delta/2$.
Thus, as the upper velocity limit is approached ($V \to \sqrt{\beta}$), the superkink becomes infinitely wide ($z \to \infty$), while its amplitude $w_{-}-w_{+}$ tends to $\delta$. This is illustrated in Fig.~\ref{fig:strains}(c). Therefore, in contrast to the case of smooth $f(w)$ \cite{VT24}, where solutions delocalize to a constant value (a kink of zero amplitude) at the bifurcation point, here they approach a kink of finite amplitude but infinite width. Thus, in the trilinear case the bifurcation is degenerate.

The lower velocity limits are different for $\alpha>1$ and $0 \leq \alpha \leq 1$. In the case $\alpha>1$ \eqref{eq:V_limits} yields $\sqrt{\alpha}<V<\sqrt{\beta}$. In the limit $V \to \sqrt{\alpha}$ the half-width $z$ of the transition region approaches a finite positive value, as illustrated by the blue curve in Fig.~\ref{fig:vary_alpha}(a):
\beq
z \to \dfrac{\sqrt{\alpha}}{4\sqrt{3}\sqrt{\beta-\alpha}}\arctan\sqrt{\dfrac{\alpha-1}{\beta-\alpha}} \quad \text{as $V \to \sqrt{\alpha}$}.
\label{eq:z_lower_limit_alpha_above_1}
\eeq
The limiting strain $w_+$ is finite in the limit (see the blue curve in Fig.~\ref{fig:vary_alpha}(b)), while $w_{-}$ tends to infinity:
\[
w_{+} \to w_c+\dfrac{\delta (1+\alpha-2\beta)}{2(\alpha-1)}=\dfrac{\alpha b}{\alpha-1}, \quad w_{-} \approx \dfrac{\delta\sqrt{(\beta-\alpha)(\beta-1)}}{\sqrt{\alpha-1}\sqrt{V^2-\alpha}} \quad \text{as $V \to \sqrt{\alpha}$},
\]
where we used \eqref{eq:b} in the first limit. See Fig.~\ref{fig:strains}(a) for an example of a superkink near the lower velocity limit.

In the case $0 \leq \alpha \leq 1$ we have $1<V<\sqrt{\beta}$. As the lower limit of unit velocity is approached, we have
\beq
z \to \dfrac{1}{4\sqrt{3}\sqrt{\beta-1}}\arctan\sqrt{\dfrac{1-\alpha}{\beta-1}} \quad \text{as $V \to 1$},
\label{eq:z_lower_limit_alpha_leq_1}
\eeq
which yields a finite positive value for $\alpha \neq 1$ (see the black and green curves in Fig.~\ref{fig:vary_alpha}(a)) and zero at $\alpha=1$ (red curve in Fig.~\ref{fig:vary_alpha}(a)). For $0 \leq \alpha<1$ $w_+$ tends to $-\infty$ (see the black and green curves in Fig.~\ref{fig:vary_alpha}(b)), while $w_{-}$ is finite in the limit:
\[
w_{+} \approx -\dfrac{\delta\sqrt{(\beta-\alpha)(\beta-1)}}{\sqrt{1-\alpha}\sqrt{V^2-1}}, \quad w_{-} \to w_c+\dfrac{\delta (1+\alpha-2\beta)}{2(\alpha-1)} \quad \text{as $V \to 1$}.
\]
In contrast, in the symmetric case $\alpha=1$, illustrated by the red curve in Fig.~\ref{fig:vary_alpha}(b), both limiting strains become infinite in the sonic limit (recall \eqref{eq:wpm_alpha1}).
Note also that the magnitude of strain grows faster in this case (inversely proportional to $V^2-1$, rather than $(V^2-1)^{1/2}$).

\subsection{Solitary waves}
We now turn to SW solutions in the QC problem \eqref{eq:TW_QC_gen2} with $f(w)$ given by  \eqref{eq:trilinear}. Recall that such solutions satisfy
\beq
w(\xi) \to w_+ \quad \text{as $\xi \to \pm \infty$}
\label{eq:BCs_SW}
\eeq
and propagate with supersonic velocities $V$ bounded by the superkink velocity $V_{SK}(w_+)$ for given $w_+$.
The waves can be tensile or compressive, depending on whether $w_+$ is in the lower ($w_+<w_c-\delta/2$) or the upper ($w_+>w_c+\delta/2$) linear regime.

\paragraph{Tensile waves.} We start with tensile solitary waves ($w(\xi)>w_+$), which arise when $w_+<w_c-\delta/2$. In this case $1<V<V_{SK}(w_+)$, where
\beq
V_{SK}(w_+)=\sqrt{1 + \dfrac{4(\beta-\alpha)(\beta-1)\delta^2}{(1 - \alpha) (2(w_c - w_+)-\delta)^2 +8(\beta-\alpha) \delta (w_c - w_+)}}
\label{eq:Vsk1}
\eeq
is obtained by solving \eqref{eq:wplus} for $V=V_{SK}$ as a function of $w_+$. In the symmetric case $\alpha=1$ we have $b=-(\beta-1)\delta$, and \eqref{eq:Vsk1} has the much simpler form
\[
V_{SK}(w_+)=\sqrt{1-\dfrac{b}{2(w_c-w_+)}}.
\]

There are two velocity regimes that need to be considered. In the first regime, we have $1<V \leq V_{cr}(w_+)$, where $V_{cr}(w_+)$ is the critical velocity value such that $w(\xi) \leq w_c+\delta/2$ for velocities below it, i.e., the solitary wave remains confined to the first and second (intermediate) linear regimes, switching from one to another at $\xi=\pm z_1$, where $z_1>0$ depends on $V$. Continuity of $w(\xi)$ and $w'(\xi)$ and the switch conditions $w(\pm z_1)=w_c-\delta/2$ then yield
\beq
w(\xi)=
\begin{cases} w_{+}+(w_c-\delta/2-w_{+})e^{-r(|\xi|-z_1)}, & |\xi| \geq z_1,\\
w_S+\dfrac{\sqrt{(\beta-1)(V^2-1)}}{\beta-V^2}(w_c-\delta/2-w_+)\cos(q\xi), & |\xi| \leq z_1,
\end{cases}
\label{eq:TSW1}
\eeq
where $w_S$ is given by \eqref{eq:wS}, and
\beq
z_1=\dfrac{1}{q}\left(\pi-\text{arctan}\dfrac{q}{r}\right).
\label{eq:z1_TSW1}
\eeq
Setting $w(0)=w_c+\delta/2$, we obtain the upper velocity bound for this regime:
\beq
V_{cr}(w_+)=\sqrt{1+\dfrac{\delta^2(\beta-1)}{(w_c-w_{+}+\delta/2)^2}}.
\label{eq:Vcr_TSW}
\eeq

\begin{figure}
\centering
\includegraphics[width=\textwidth]{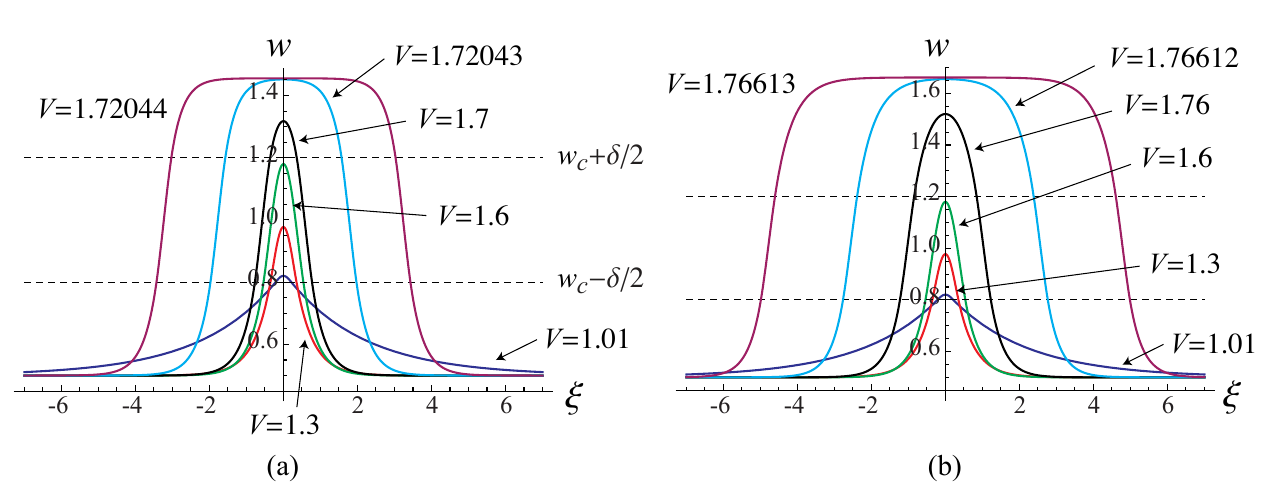}
\caption{\footnotesize Tensile solitary waves in the trilinear model \eqref{eq:trilinear} at different velocities at $w_+=0.5$ and
(a) $\alpha=0.5$; (b) $\alpha=2$. Here $\beta=6$, $w_c=1$, $\delta=0.4$.}
\label{fig:TSW_trilinear}
\end{figure}

For $V_{cr}(w_+)< V<V_{SK}(w_+)$, the solution involves all three linear regions and has the form
\beq
w(\xi)=\begin{cases} w_{+}+Ae^{-r(|\xi|-z_1)}, & |\xi| \geq z_1,\\
w_S+B\cos(q\xi)+C\sin(q|\xi|), & z_2 \leq |\xi| \leq z_1,\\
w_{-}+D\cosh(s\xi), & |\xi| \leq z_2,
\end{cases}
\label{eq:TSW2}
\eeq
where the coefficients $A$, $B$, $C$, $D$ listed in \eqref{eq:TSW2_coefs} in Appendix~\ref{app:coefs} are found by imposing the continuity of the strain and its derivative, $w_{-}$ is given by \eqref{eq:RH} for $\alpha>0$ and by \eqref{eq:RH_alpha0} for $\alpha=0$, and $w_S$ is given by \eqref{eq:wS}.
Imposing $w(\pm z_1)=w_c-\delta/2$ and $w(\pm z_2)=w_c+\delta/2$ then yields
\beq
z_2=\dfrac{1}{s}\text{arctanh}\dfrac{\sqrt{(V^2-1)(w_c-w_{+}+\delta/2)^2-(\beta-1)\delta^2}}{\sqrt{V^2-\alpha}(w_{-}-w_c-\delta/2)}
\label{eq:z2_TSW2}
\eeq
and
\beq
z_1=z_2+\dfrac{1}{q}\left(\text{arccos}\dfrac{(V^2-1)(w_c+\delta/2-w_{+})-(\beta-1)\delta}{\sqrt{(\beta-1)(V^2-1)}(w_c-\delta/2-w_{+})}
-\text{arctan}\dfrac{q}{r}\right).
\label{eq:z1_TSW2}
\eeq
Note that $z_2=0$ at $V=V_{cr}(w_+)$ and that $z_2 \to \infty$ as $V \to V_{SK}(w_+)$, which means that the width of the solitary waves tends to infinity as the upper velocity limit is approached. One can also verify that in this limit $w(0) \to w_{-}$ and $z_1-z_2 \to 2z$, where we recall that $z$, given by \eqref{eq:z} for $\alpha \neq 1$ and \eqref{eq:z_alpha1} for $\alpha=1$, is the half-width of the transition interval for a superkink solution. This is consistent with the SW solution approaching a kink-antikink bundle, as discussed in \cite{Gorbushin19,Gorbushin21,VT24}, at velocities just below the superkink limit. Examples of tensile solitary waves illustrating this are shown in Fig.~\ref{fig:TSW_trilinear}.
Observe also that the near-sonic solutions, which are given by \eqref{eq:TSW1} and are independent of $\alpha$, delocalize to the constant strain $w(\xi)=w_c-\delta/2$ as $V \to 1$ because $r \to 0$ in \eqref{eq:roots} tends to zero in this limit.

\paragraph{Compressive waves.} Similarly, we can obtain compressive SW solutions, which arise when $w_+>w_c+\delta/2$ and $\sqrt{\alpha}<V<V_{SK}(w_+)$, where
\beq
V_{SK}(w_+)=\sqrt{\alpha + \dfrac{4(\beta-\alpha)(\beta-1)\delta^2}{(\alpha-1) (2(w_+ - w_c)-\delta)^2 +8(\beta-1) \delta (w_+ - w_c)}}
\label{eq:Vsk2}
\eeq
is obtained by setting the right hand side of \eqref{eq:wminus} equal to $w_+$ and solving the resulting equation for $V=V_{SK}$ as a function of $w_+$. The superkink velocity simplifies to
\[
V_{SK}(w_+)=\sqrt{1-\dfrac{b}{2(w_{+}-w_c)}}
\]
in the case $\alpha=1$.

\begin{figure}[h]
\centering
\includegraphics[width=\textwidth]{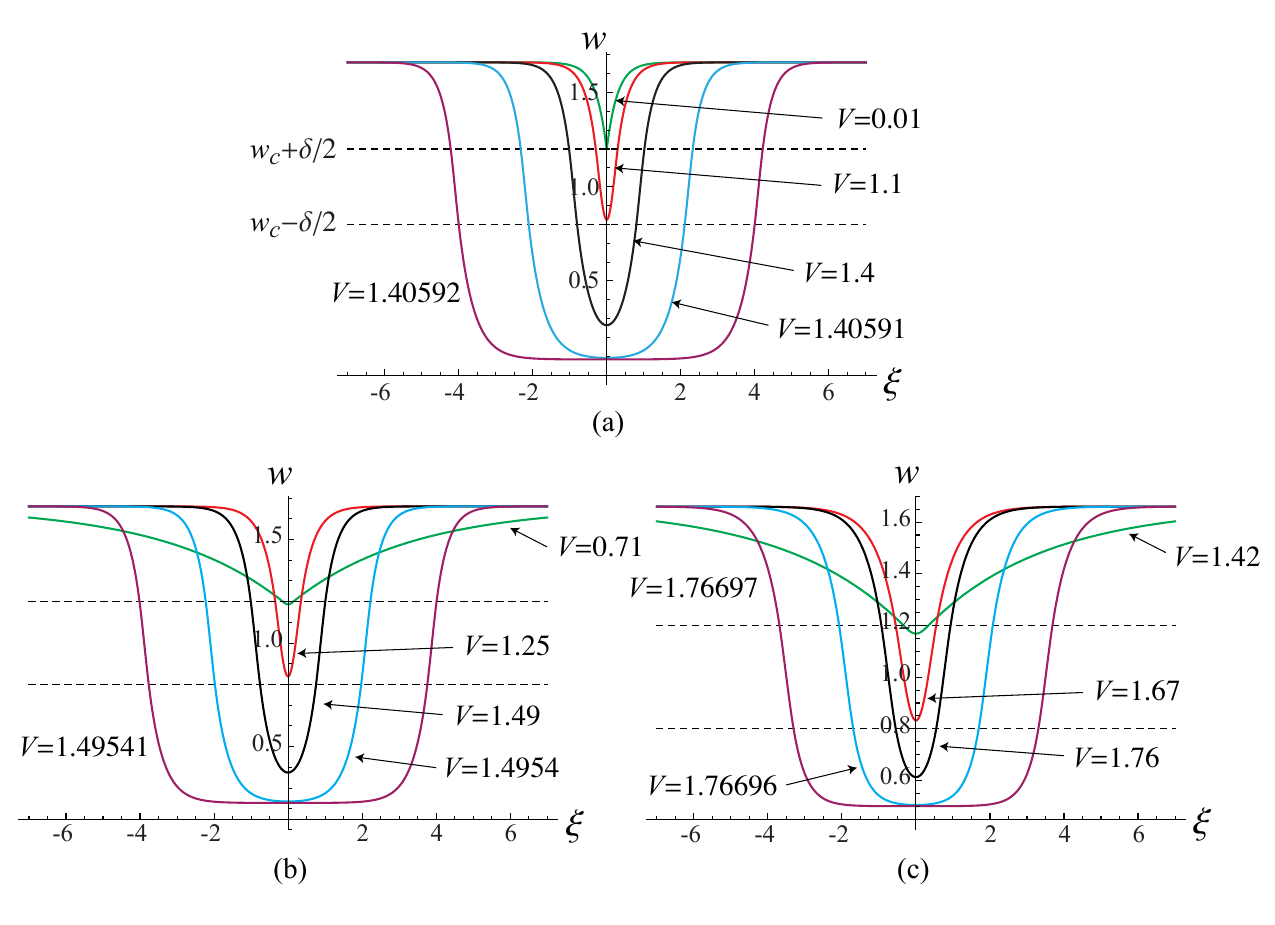}
\caption{\footnotesize Compressive solitary waves in the trilinear model \eqref{eq:trilinear} at different velocities at $w_+=1.66$ and
(a) $\alpha=0$; (b) $\alpha=0.5$; (c) $\alpha=2$. Here $\beta=6$, $w_c=1$, $\delta=0.4$.}
\label{fig:CSW_trilinear}
\end{figure}

In this case we also have two velocity regimes. In the first, $\sqrt{\alpha}<V \leq V_{cr}(w_+)$, where
\beq
V_{cr}(w_+)=\sqrt{\alpha+\dfrac{\delta^2(\beta-\alpha)}{(w_{+}-w_c+\delta/2)^2}},
\label{eq:Vcr_CSW}
\eeq
and we have
\beq
w(\xi)=
\begin{cases} w_{+}-(w_{+}-w_c-\delta/2)e^{-s(|\xi|-z_1)}, & |\xi| \geq z_1,\\
w_S-\dfrac{\sqrt{(\beta-\alpha)(V^2-\alpha)}}{\beta-V^2}(w_{+}-w_c-\delta/2)\cos(q\xi), & |\xi| \leq z_1,
\end{cases}
\label{eq:CSW1}
\eeq
with
\beq
w_S=\dfrac{(\beta-1)(w_c-\delta/2)-(V^2-1)w_{-}}{\beta-V^2},
\label{eq:wS_CSW}
\eeq
\beq
z_1=\dfrac{1}{q}\left(\pi-\text{arctan}\dfrac{q}{s}\right)
\label{eq:z1_CSW1}
\eeq
and
\[
w_{-}=\dfrac{(V^2-\alpha)w_{+}+\alpha b}{V^2-1}.
\]
In the second regime, $V_{cr}(w_+)<V<V_{SK}(w_+)$, the solution has the form
\beq
w(\xi)=\begin{cases} w_{+}+Ae^{-s(|\xi|-z_1)}, & |\xi| \geq z_1,\\
w_S+B\cos(q\xi)+C\sin(q|\xi|), & z_2 \leq |\xi| \leq z_1,\\
w_{-}+D\cosh(r\xi), & |\xi| \leq z_2,
\end{cases}
\label{eq:CSW2}
\eeq
where $w_S$ is provided in \eqref{eq:wS_CSW},
the coefficients $A$, $B$, $C$ and $D$ are given by \eqref{eq:CSW2_coefs},
and we have
\beq
z_2=\dfrac{1}{r}\text{arctanh}\dfrac{\sqrt{(V^2-\alpha)(w_{+}-w_c+\delta/2)^2-(\beta-\alpha)\delta^2}}{\sqrt{V^2-1}(w_c-\delta/2-w_{-})}
\label{eq:z2_CSW2}
\eeq
and
\beq
z_1=z_2+\dfrac{1}{q}\left(\pi-\text{arctan}\dfrac{q}{s}-\text{arccos}\dfrac{(V^2-1)(w_c-\delta/2-w_{-})}
{\sqrt{(\beta-\alpha)(V^2-\alpha)}(w_{+}-w_c-\delta/2)}\right).
\label{eq:z1_CSW2}
\eeq
Similar to the tensile SW solutions, we have $z_2 \to \infty$, $z_1-z_2 \to 2z$ and $w(0) \to w_{-}$ in the limit $V \to V_{SK}(w_+)$, with solutions just below the limit have the kink-antikink structure, as illustrated in Fig.~\ref{fig:CSW_trilinear}.

{\paragraph{Sonic limit and the singular $\alpha=0$ case.} In the case of compressive solitary waves the near-sonic behavior depends on $\alpha$. When $\alpha>0$, solutions \eqref{eq:CSW1} delocalize to the constant strain $w(\xi)=w_c+\delta/2$ in the sonic limit, Indeed, observe that $s \to 0$ as $V \to \sqrt{\alpha}$, so that $w(\xi) \to w_c+\delta/2$ for $|\xi|\geq z_1$, while $(V^2-1)w_{-} \to \alpha b$, which together with \eqref{eq:b} yields the limit $w(\xi) \to w_c+\delta/2$ for $|\xi|\leq z_1$. Note that in this case $z_1$ in \eqref{eq:z1_CSW1} is nonzero in the sonic limit.

However, when $\alpha=0$, the exponent $s=\sqrt{12}$ in \eqref{eq:CSW1} is independent of $V$. Note also that in this case $z_1 \to 0$ in \eqref{eq:z1_CSW1} as $V \to 0$ because $q$ in \eqref{eq:roots} tends to infinity in this limit. Thus, in this case we have
a \emph{nontrivial sonic limit}
\beq
w(x) \to w_{+}-(w_{+}-w_c-\delta/2)\exp(-\sqrt{12}|x|),
\label{eq:sonic_limit}
\eeq
as $V \to 0$, and thus $\xi=x-Vt \to x$.
This is a manifestation of the fact that in the case $\alpha=0$ the sound speed is zero, and slowly moving solitary waves are effectively replacing the conventional linear elastic waves as a elementary ``quanta'' of mechanical information. Similar effects were first discovered in Hertzian granular chains without precompression \cite{Nesterenko01,Sen08,Chong17}.

\section{Traveling wave solutions of the discrete problem}
\label{sec:discrete}
We now turn our attention to traveling wave solutions of the discrete problem \eqref{eq:FPUstrain}. As in the QC case, it suffices to consider $V>0$.
To compute these solutions, we follow the approach in \cite{Aubry09,Vainchtein20,James21,VT24} that exploits the fact that traveling
waves satisfy \eqref{eq:TWansatz} and hence are periodic modulo shift:
\beq
w_{n+1}(t+T)=w_n(t), \quad T=1/V.
\label{eq:period}
\eeq
This implies that such solutions are fixed points
of the nonlinear map
\begin{equation}
\label{eq:nonlin_map}
\left[\begin{array}{c}
  \{w_{n+1}(T)\} \\ \{\dot w_{n+1}(T)\} \\  \end{array}\right]
  =\mathcal{N}\left(
  \left[\begin{array}{c}
  \{w_{n}(0)\} \\ \{\dot w_{n}(0)\} \\  \end{array}\right]\right),
\end{equation}
where $\mathcal{N}$ is defined by integrating \eqref{eq:FPUstrain} over one period for given initial data over the period $T$ and then shifting the obtained solution by one lattice space. This nonlinear map approach dates back to computation of discrete breathers \cite{MarinAubry96}.

\subsection{Superkinks}
To compute the superkink solutions of the discrete problem \eqref{eq:FPUstrain} propagating with given velocity $V$, we use Newton-Raphson iteration procedure with finite-difference Jacobian to solve
\beq
\begin{split}
&w_{n+1}(T)=w_n(0), \quad n=-N/2,\dots,N/2-1, \\
&\dot{w}_{n+1}(T)=\dot{w}_n(0), \quad n=-N/2,\dots,N/2-2, \quad w_1(T)=w_*,
\end{split}
\label{eq:nonlin_system}
\eeq
where $T=1/V$ and $N \geq 400$ is an even number, for $\{w_n(0),\dot{w}_n(0)\}$, $n=-N/2,\dots,N/2-1$. At each iteration, $w_n(T)$ and $\dot{w}_n(T)$ are obtained for given $w_n(0)$ and $\dot{w}_n(0)$ by using Dormand-Prince algorithm (Matlab's ode45 routine) to integrate \eqref{eq:FPUstrain} on the finite chain with the boundary conditions
\[
w_{-N/2-1}(t)=w_{-}, \quad w_{N/2}(t)=w_{+},
\]
where $w_{\pm}$ are given by \eqref{eq:wplus}, \eqref{eq:wminus} for $0 \leq \alpha \neq 1$ and \eqref{eq:wpm_alpha1} for $\alpha=1$. The last equation in \eqref{eq:nonlin_system} is a pinning condition, which is necessary to eliminate the non-uniqueness due to the time-translational invariance. To enable the comparison with superkink solutions $w_{QC}(\xi)$ of the QC model, which are also used to obtain an initial guess for the Newton-Raphson procedure, we set $w_*=w_{QC}(0)$, so that $w_0(0)=w_1(T)=w_{QC}(0)$ and thus the traveling wave $w(\xi)$ for the discrete problem satisfies $w(0)=w_{QC}(0)$. To obtain a system of $2N$ nonlinear equations for $2N$ unknowns while prescribing the pinning condition, we drop the equation for $\dot{w}_{N/2}(T)$ in \eqref{eq:nonlin_system}. Due to the large size of the computational domain, the equation is automatically satisfied for the computed solutions within the numerical tolerance of $10^{-13}$.

The computed strain profiles $w_n(0)=w(n)$ for are shown in Fig.~\ref{fig:TWtrilinear} together with the corresponding profiles $w(x)$ obtained from the exact solutions of the QC model.
\begin{figure}
\centering
\includegraphics[width=\textwidth]{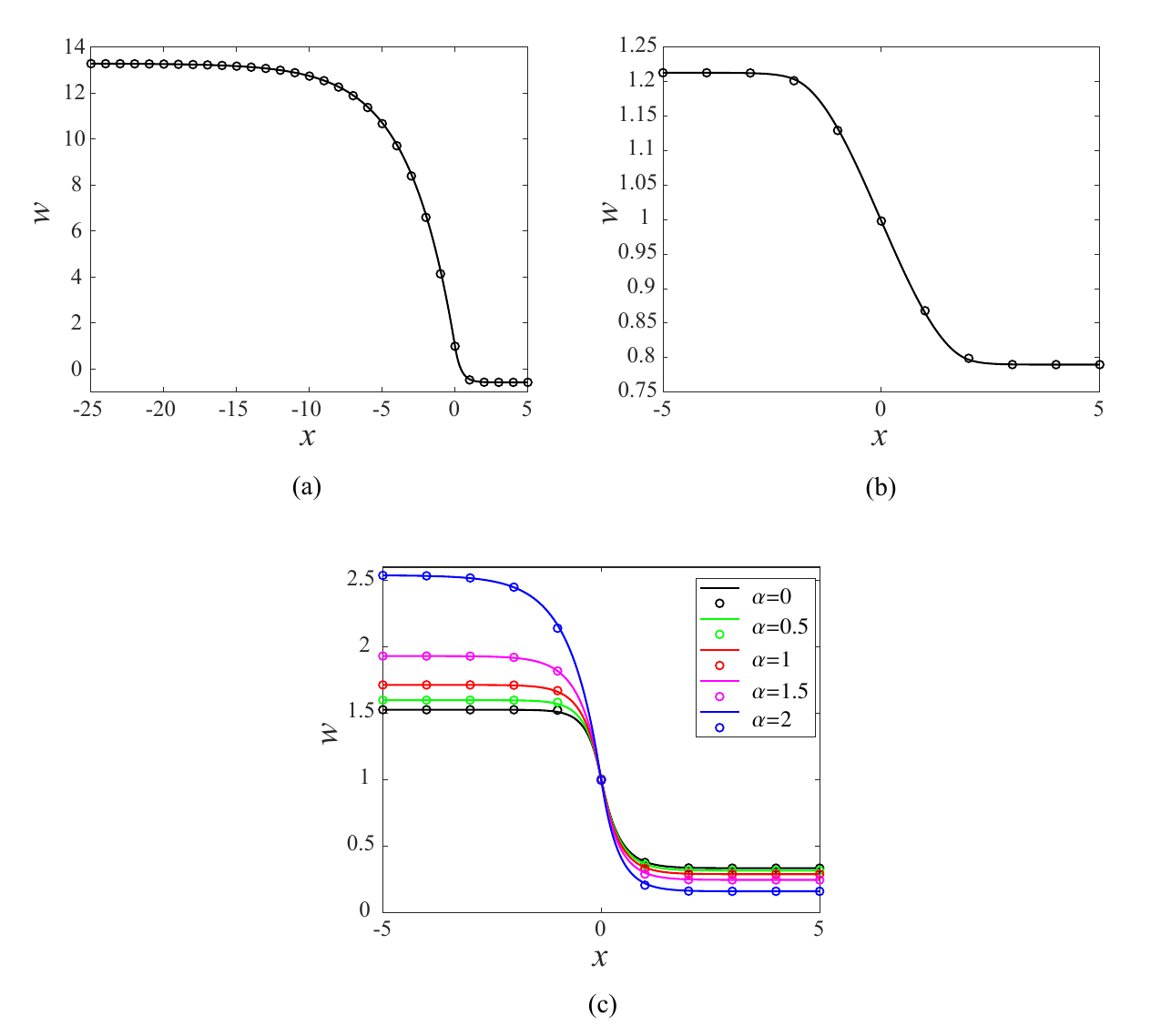}
\caption{\footnotesize Superkink solutions $w_n(0)=w(n)$ of the discrete problem \eqref{eq:TW_discrete} (circles) with trilinear $f(w)$ and the corresponding solutions $w(x)$ for the QC model \eqref{eq:TW_QC_gen2} (solid curves) evaluated at $t=0$ for (a) $V=1.42$, $\alpha=2$; (b) $V=2.4$, $\alpha=2$; (c) $V=1.55$ at different values of $\alpha$, as indicated in the legend. Here $\beta=6$, $\delta=0.4$, $w_c=1$.}
\label{fig:TWtrilinear}
\end{figure}
One can see that in the cases shown superkink solutions there is a very good agreement between the discrete and QC models, particularly near the sonic limits (panels (a) and (b)), where the solutions largely involve the long-wave contributions that are well captured by the QC model. Closer inspection of panel (c), however, reveals some discrepancies between solutions of the two models inside the transition layer ($x=\pm 1$).

\subsection{Solitary waves}
To compute SW solutions with prescribed far-field strain $w_+$ and velocity $V$, we use the approach we employed in the case of superkinks, except that in this case the boundary conditions are $w_{-N/2-1}(t)=w_{N/2}(t)=w_{+}$, and the pinning condition is $\dot{w}_1(T)=0$. The latter ensures that the maximum of a tensile solitary wave (or the minimum of a compressive one) is at $n=0$ when $t=0$. We use parameter continuation in $V$ to compute solutions in the entire velocity range.

The results for tensile waves at $\alpha=0.5$ and $\alpha=2$ are shown in Fig.~\ref{fig:triTSW_DvsQC}-\ref{fig:triTSWenergy_DvsQC}, while the corresponding results for the compressive waves
can be seen in Fig.~\ref{fig:triCSW_DvsQC}-\ref{fig:triCSWenergy_DvsQC}. In addition to direct comparison of the discrete and QC solitary waves in Fig.~\ref{fig:triTSW_DvsQC} and Fig.~\ref{fig:triCSW_DvsQC}, we show amplitude-velocity plots in Fig.~\ref{fig:triTSWamp_DvsQC} and Fig.~\ref{fig:triCSWamp_DvsQC}, as well as energy-velocity plots in Fig.~\ref{fig:triTSWenergy_DvsQC} and Fig.~\ref{fig:triCSWenergy_DvsQC}. Since the energy of the waves with nonzero background is infinite, we renormalize it by subtracting the energy of the background, as in \cite{VT24}. For the discrete model this yields
\beq
E^D_{ren}(V)=\sum_n \left\{\dfrac{1}{2}v_n^2+\dfrac{1}{2}\left(\Phi(w_n)+\Phi(w_{n+1})\right)-\Phi(w_+)-\dfrac{1}{2}V^2 w_+^2\right\}
\label{eq:Eren_D}
\eeq
in the discrete case, where $v_n$ are the particle velocities, and all values are evaluated at $t=0$ due to the energy conservation.
For the QC model we have
\beq
E^{QC}_{ren}(V)=\int_{-\infty}^{\infty} \left\{\dfrac{1}{2}V^2 w^2(\xi)+\dfrac{1}{24}V^2 (w'(\xi))^2+\Phi(w(\xi))-\Phi(w_+)-\dfrac{1}{2}V^2 w_+^2\right\}d\xi,
\label{eq:Eren_QC}
\eeq
where we used the fact that for a traveling wave solution $w(\xi)=w(x-Vt)$ we have $v(\xi)=-V w(\xi)$.

One can see that the discrepancy between discrete and QC solutions is much more pronounced in the case of solitary waves. For tensile waves, the disagreement between the two models is greater for $\alpha=0.5$, while for compressive waves there is more discrepancy at $\alpha=2$. In addition to overstimating the amplitude of solitary waves away from the superkink limit, the QC model does not correctly capture their width. Note that the width in the QC case is controlled by $z_2$ in \eqref{eq:z2_TSW2} and \eqref{eq:z2_CSW2}, which involve approximations of the relevant $\alpha$-dependent roots of the discrete problem that become worse at velocities away from $\sqrt{\alpha}$. Nevertheless, the QC model captures the evolution of the SW solutions of the discrete problem qualitatively, and one can see quantitative agreement near the sonic and superkink limits.
\begin{figure}
\centering
\includegraphics[width=\textwidth]{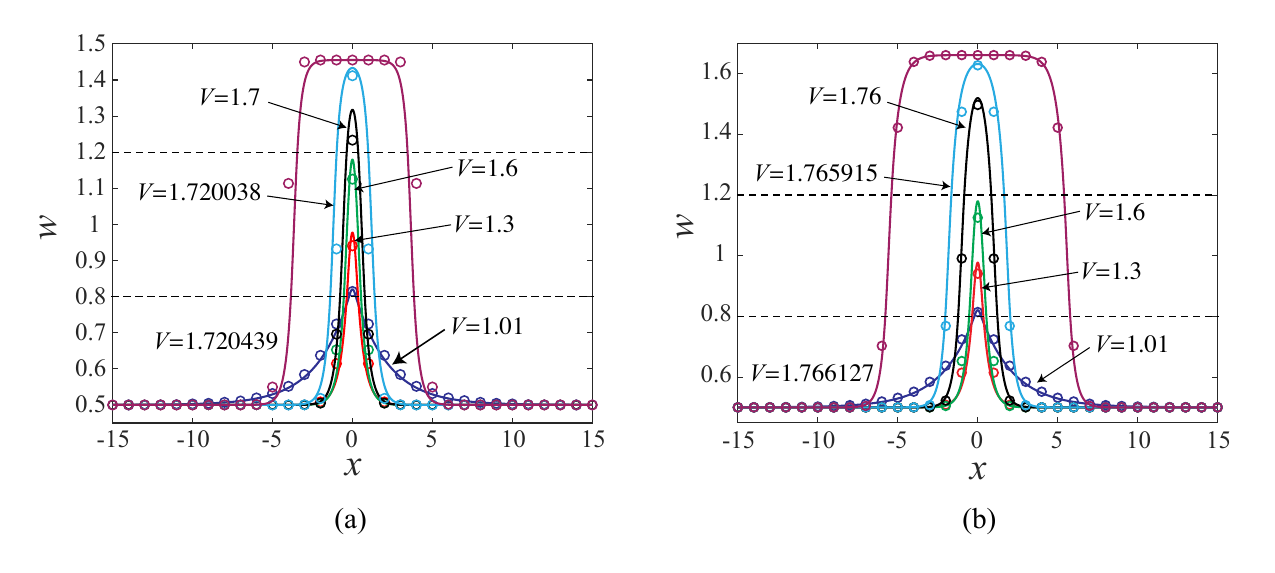}
\caption{\footnotesize Tensile SW solutions $w_n(0)=w(n)$ of the discrete problem \eqref{eq:TW_discrete} (circles) with trilinear $f(w)$ and the corresponding solutions $w(x)$ for the QC model \eqref{eq:TW_QC_gen2} (solid curves) evaluated at $t=0$ for (a) $\alpha=0.5$; (b) $\alpha=2$. Here $\beta=6$, $\delta=0.4$, $w_c=1$. The background strain is $w_+=0.5$, and the corresponding superkink velocity is $V_{SK}=1.72044$ in (a) and $V_{SK}=1.76613$ in (b).}
\label{fig:triTSW_DvsQC}
\end{figure}
\begin{figure}
\centering
\includegraphics[width=\textwidth]{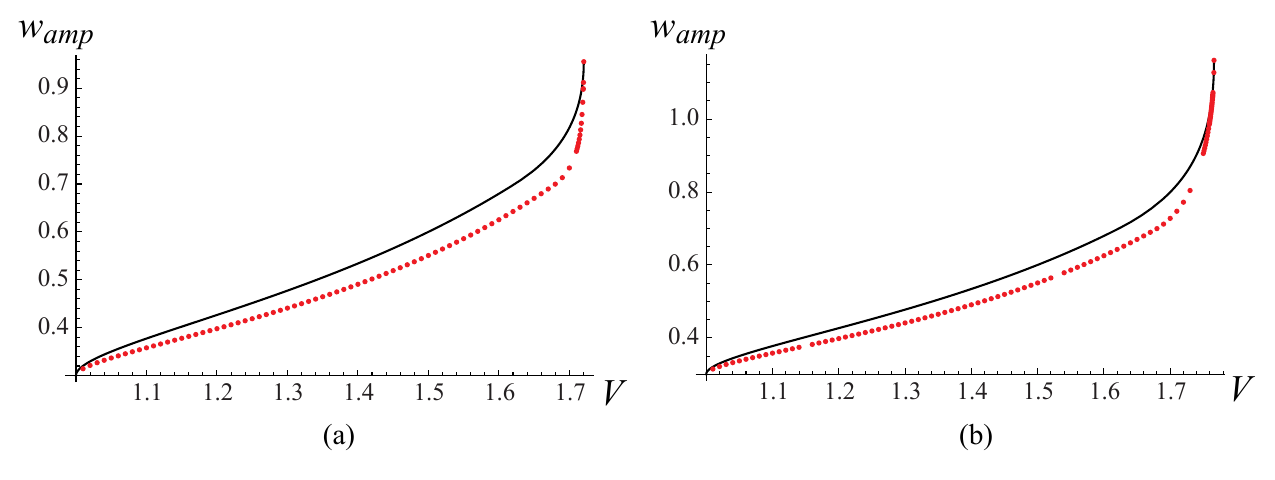}
\caption{\footnotesize Amplitude $w_{amp}=|w(0)-w_+|$ as a function of velocity $V$ for tensile SW solutions of the discrete problem \eqref{eq:TW_discrete} (circles) with trilinear $f(w)$ and the corresponding solutions for the QC model \eqref{eq:TW_QC_gen2} (solid curves) evaluated for (a) $\alpha=0.5$; (b) $\alpha=2$. Here $\beta=6$, $\delta=0.4$, $w_c=1$. The background strain is $w_+=0.5$, and the corresponding superkink velocity is $V_{SK}=1.72044$ in (a) and $V_{SK}=1.76613$ in (b).}
\label{fig:triTSWamp_DvsQC}
\end{figure}
\begin{figure}
\centering
\includegraphics[width=\textwidth]{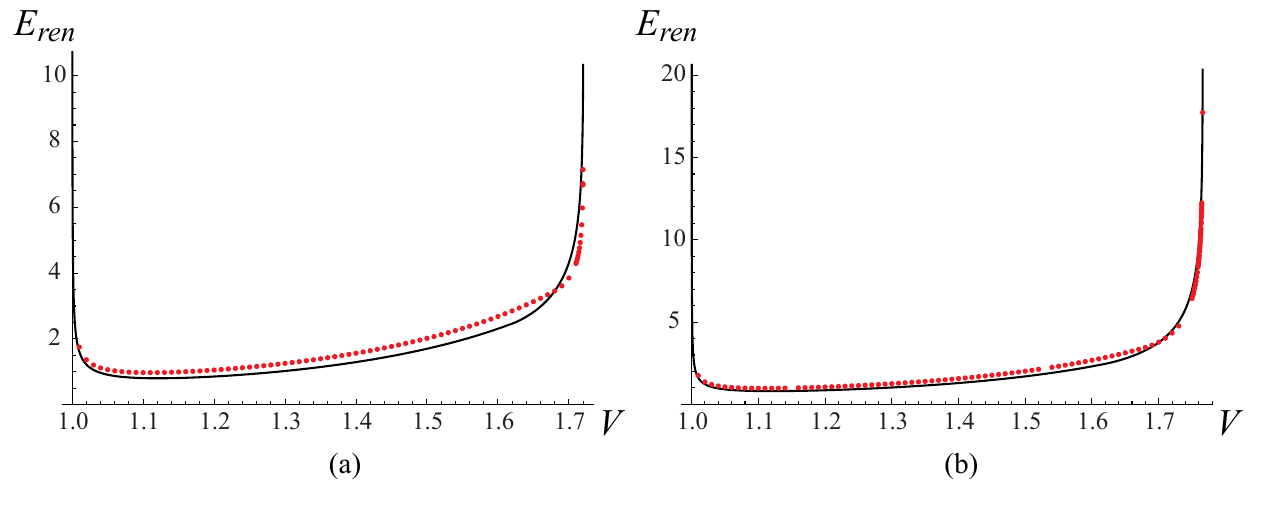}
\caption{\footnotesize Renormalized energy $E_{ren}$ given by \eqref{eq:Eren_D} as a function of velocity $V$ for tensile SW solutions of the discrete problem \eqref{eq:TW_discrete} (circles) with trilinear $f(w)$ and the corresponding energy \eqref{eq:Eren_QC} for the QC model \eqref{eq:TW_QC_gen2} (solid curves) evaluated for (a) $\alpha=0.5$; (b) $\alpha=2$. Here $\beta=6$, $\delta=0.4$, $w_c=1$. The background strain is $w_+=0.5$, and the corresponding superkink velocity is $V_{SK}=1.72044$ in (a) and $V_{SK}=1.76613$ in (b).}
\label{fig:triTSWenergy_DvsQC}
\end{figure}
\begin{figure}
\centering
\includegraphics[width=\textwidth]{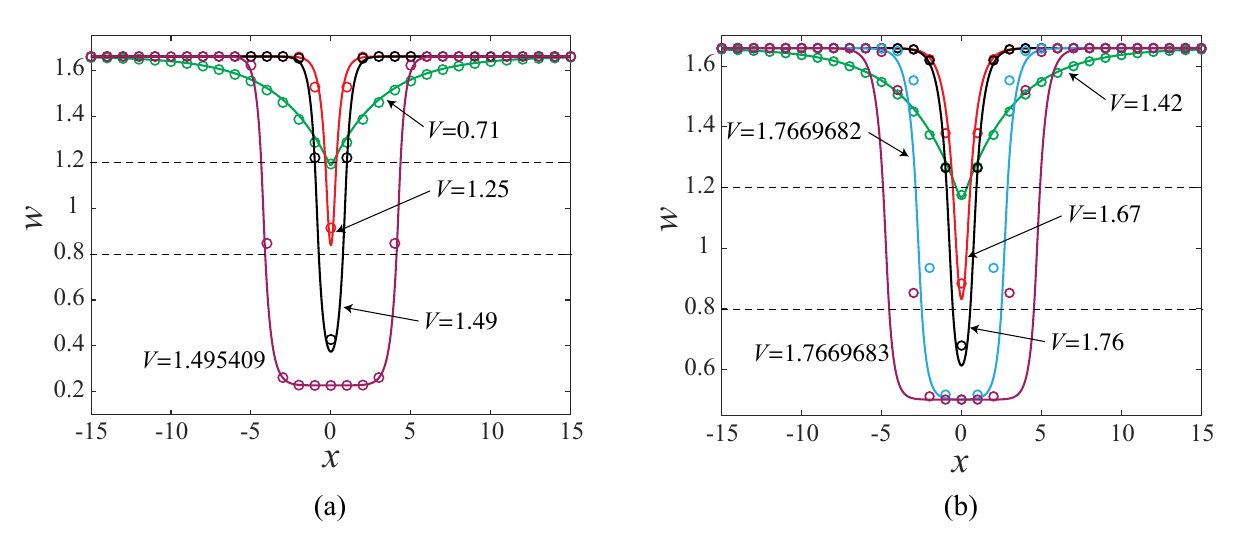}
\caption{\footnotesize Compressive SW solutions $w_n(0)=w(n)$ of the discrete problem \eqref{eq:TW_discrete} (circles) with trilinear $f(w)$ and the corresponding solutions $w(x)$ for the QC model \eqref{eq:TW_QC_gen2} (solid curves) evaluated at $t=0$ for (a) $\alpha=0.5$; (b) $\alpha=2$. Here $\beta=6$, $\delta=0.4$, $w_c=1$. The background strain is $w_+=1.66$, and the corresponding superkink velocity is $V_{SK}=1.49541$ in (a) and $V_{SK}=1.76697$ in (b).}
\label{fig:triCSW_DvsQC}
\end{figure}
\begin{figure}
\centering
\includegraphics[width=\textwidth]{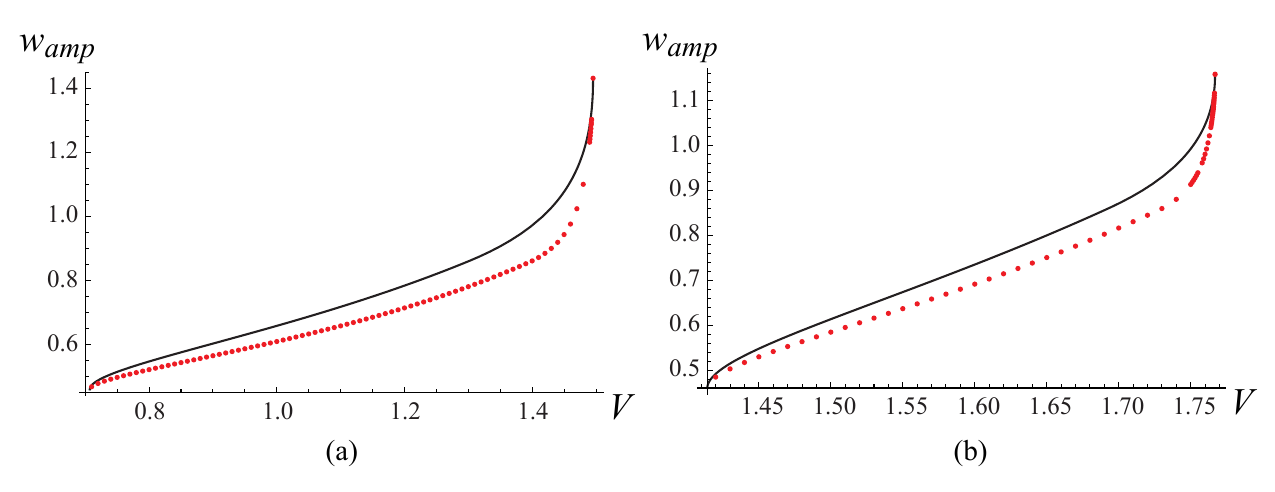}
\caption{\footnotesize Amplitude as a function of velocity $V$ for compressive SW solutions of the discrete problem \eqref{eq:TW_discrete} (circles) with trilinear $f(w)$ and the corresponding solutions for the QC model \eqref{eq:TW_QC_gen2} (solid curves) for (a) $\alpha=0.5$; (b) $\alpha=2$. Here $\beta=6$, $\delta=0.4$, $w_c=1$. The background strain is $w_+=1.66$, and the corresponding superkink velocity is $V_{SK}=1.49541$ in (a) and $V_{SK}=1.76697$ in (b).}
\label{fig:triCSWamp_DvsQC}
\end{figure}
\begin{figure}
\centering
\includegraphics[width=\textwidth]{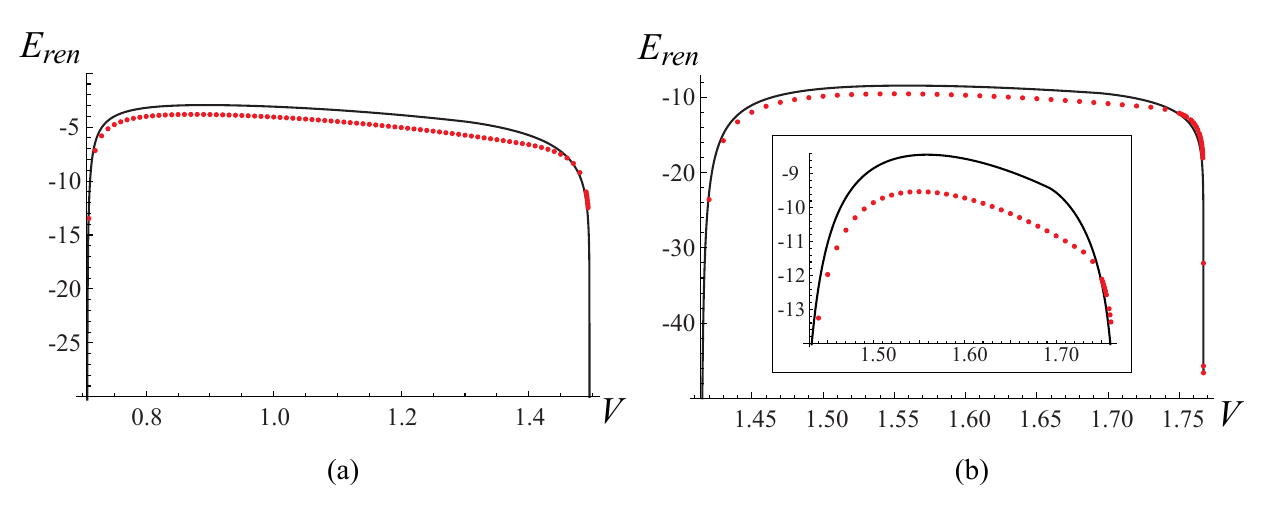}
\caption{\footnotesize Renormalized energy as a function of velocity $V$ for compressive SW solutions of the discrete problem \eqref{eq:TW_discrete} (circles) with trilinear $f(w)$ and the corresponding solutions of the QC model \eqref{eq:TW_QC_gen2} (solid curves) for (a) $\alpha=0.5$; (b) $\alpha=2$. The inset in (b) zooms in on the upper energy range. Here $\beta=6$, $\delta=0.4$, $w_c=1$. The background strain is $w_+=1.66$, and the corresponding superkink velocity is $V_{SK}=1.49541$ in (a) and $V_{SK}=1.76697$ in (b).}
\label{fig:triCSWenergy_DvsQC}
\end{figure}

\paragraph{Compressive solitary waves at $\alpha=0$.} Of particular interest are compressive solitary waves at $\alpha=0$. The results for this singular limit are shown in Fig.~\ref{fig:triCSW_DvsQC_alpha0} and Fig.~\ref{fig:triCSWampen_DvsQC_alpha0}.
\begin{figure}
\centering
\includegraphics[width=\textwidth]{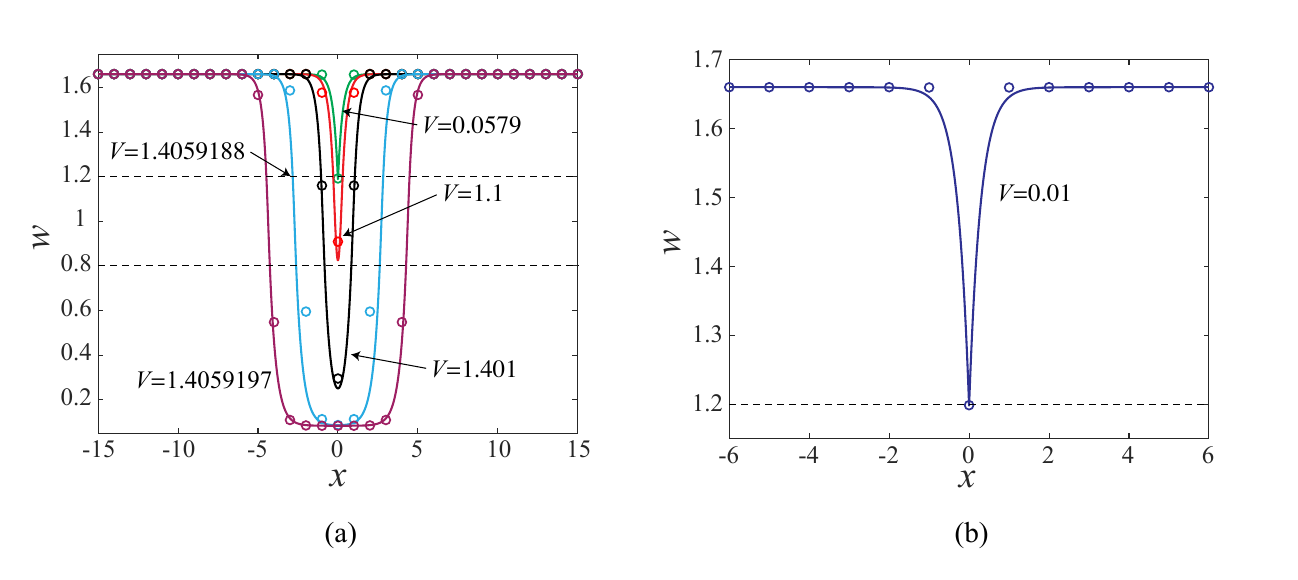}
\caption{\footnotesize (a) Compressive SW solutions $w_n(0)=w(n)$ of the discrete problem \eqref{eq:TW_discrete} (circles) with trilinear $f(w)$ and the corresponding solutions $w(x)$ for the QC model \eqref{eq:TW_QC_gen2} (solid curves) evaluated at $t=0$ for $\alpha=0$, $\beta=6$, $\delta=0.4$, $w_c=1$. (b) The solutions at $V=0.01$. The background strain is $w_+=1.66$, and the corresponding superkink velocity is $V_{SK}=1.40592$.}
\label{fig:triCSW_DvsQC_alpha0}
\end{figure}
\begin{figure}
\centering
\includegraphics[width=\textwidth]{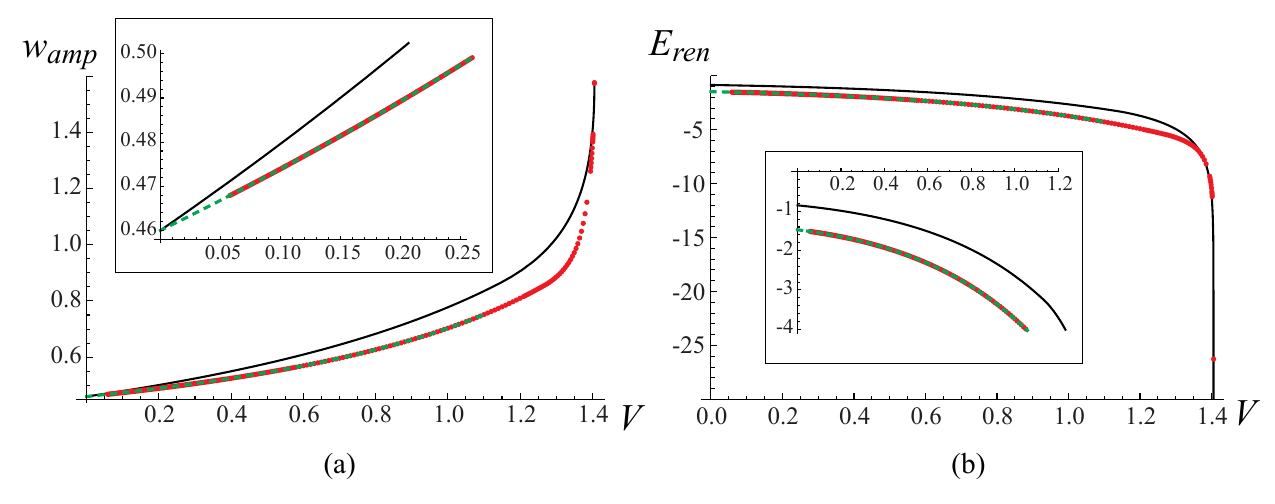}
\caption{\footnotesize (a) Amplitude and (b) renormalized energy as functions of velocity $V$ for compressive SW solutions of the discrete problem \eqref{eq:TW_discrete} (circles) with trilinear $f(w)$ and the corresponding solutions for the QC model \eqref{eq:TW_QC_gen2} (solid curves) for $\alpha=0$. The dashed green curves correspond to the exact small-velocity solution \eqref{eq:exact_CSW_1}-\eqref{eq:exact_CSW_3}. Insets zoom in around smaller velocity values. Here $\beta=6$, $\delta=0.4$, $w_c=1$. The background strain is $w_+=1.66$, and the corresponding superkink velocity is $V_{SK}=1.40592$.}
\label{fig:triCSWampen_DvsQC_alpha0}
\end{figure}

In this case we can obtain exact solutions for small enough velocities. By periodicity modulo shift it suffices to consider the time interval $[0,T]$, where we recall that $T=1/V$. Suppose at $t=0$ the strain $w_n$ has even symmetry about $n=0$, and $|w_0-w_c|<\delta/2$, while all other strains satisfy $w_n>w_c+\delta/2$. Let $t=T_1$ be the time when $w_0(t)$ switches to the (degenerate) upper linear regime, $w_0(T_1)=w_c+\delta/2$. Then by symmetry $w_1(t)$ switches to the intermediate linear regime at $t=T_2=T-T_1$. In what follows, we assume that $T_1<T_2$, i.e., $T_1<T/2$. Under these assumptions we find that for $0 \leq t<T_1$ \eqref{eq:FPUstrain} with \eqref{eq:trilinear} at $\alpha=0$ reduce to
\[
\ddot{w}_0+2\beta w_0=\beta(2 w_c+\delta), \quad \ddot{w}_{\pm 1}=\beta(w_0(t)-w_c)-\beta\delta/2, \quad \ddot{w}_n=0,
\quad |n| \geq 2.
\]
Solving these and imposing the symmetry conditions $\ddot{w}_0(0)=0$, $w_1(0)=w_{-1}(0)$, $\dot{w}_1(0)=-\dot{w}_{-1}(0)$ and the boundary condition $w_n \to w_+$ as $|n| \to \infty$, we obtain
\beq
\begin{split}
w_0&=A\cos(\sqrt{2\beta}t)+w_c+\delta/2, \quad w_{\pm 1}= -(A/2)\cos(\sqrt{2\beta}t)\pm Bt+C, \\
w_n&=w_+,\quad n \leq -2,\quad n \geq 2
\end{split}
\label{eq:exact_CSW_1}
\eeq
for $t \leq T_1$. The condition $w_0(T_1)=w_c+\delta/2$ then yields $A\cos(\sqrt{2\beta} T_1)=0$, and since $A$ must be nonzero, we deduce that $\sqrt{2\beta} T_1=\pi/2$, which yields
\[
T_1=\dfrac{\pi}{2\sqrt{2\beta}}.
\]
Since $T_1<T/2$, we have
\[
0<V<\dfrac{\sqrt{2\beta}}{\pi}.
\]
For $T_1<t<T_2$, all strains are in the upper linear regime and satisfy $\ddot{w}_n=0$. Together with continuity of $w_n$ and $\dot{w}_n$ at $t=T_1$ this yields
\beq
\begin{split}
w_0&=w_c+\delta/2-\sqrt{2\beta}A(t-T_1), \quad w_{\pm 1}=C\pm Bt+\sqrt{\beta/2}A(t-T_1), \\
w_n&=w_+,\quad n \leq -2,\quad n \geq 2
\end{split}
\label{eq:exact_CSW_2}
\eeq
for $T_1 \leq t \leq T-T_1$. For $T_2<t\leq T$, we have
\[
\ddot{w}_1+2\beta w_1=\beta(2 w_c+\delta), \quad \ddot{w}_{0,2}=\beta(w_1(t)-w_c)-\beta\delta/2, \quad \ddot{w}_n=0,
\quad n \leq -1,\quad n \geq 3.
\]
Solving these and recalling that $w_n(0)=w_{n+1}(T)$ and $\dot{w}_n(0)=\dot{w}_{n+1}(T)$, we obtain
\beq
\begin{split}
w_1&=A\cos(\sqrt{2\beta}(t-T))+w_c+\delta/2, \quad w_{0,2}=-(A/2)\cos(\sqrt{2\beta}(t-T))\mp B(t-T)+C,\\
w_n&=w_+,\quad n \leq -1,\quad n \geq 3
\end{split}
\label{eq:exact_CSW_3}
\eeq
for $T-T_1 \leq t \leq T$.
Continuity of $w_n$ and $\dot{w}_n$ at $t=T_2$ then leads to three independent conditions
\[
\begin{split}
&C+B(T-T_1)+\sqrt{\beta/2}A(T-2T_1)=w_c+\delta/2,\\
&C-B(T-T_1)+\sqrt{\beta/2}A(T-2T_1)=w_{+},\\
&\sqrt{\beta/2}A-B=0,
\end{split}
\]
which yield
\[
B=-\dfrac{w_{+}-w_c-\delta/2}{2(T-T_1)}=\sqrt{\dfrac{\beta}{2}}A,
\quad C=w_{+}-\dfrac{w_{+}-w_c-\delta/2}{2(T-T_1)}T_1.
\]
In particular, at $t=0$ we obtain
\beq
w_0=w_c+\dfrac{\delta}{2}-\dfrac{w_{+}-w_c-\delta/2}{\sqrt{2\beta}(T-T_1)}, \quad
w_{\pm 1}=w_{+}-\dfrac{w_{+}-w_c-\delta/2}{2\sqrt{2\beta}(T-T_1)}\left(\dfrac{\pi}{2}-1\right), \quad w_n=w_+, \quad |n| \geq 2
\label{eq:exact_CSW_strains}
\eeq
and
\[
\dot{w}_{\pm 1}=\mp \dfrac{w_{+}-w_c-\delta/2}{2(T-T_1)}, \quad \dot{w}_0=0, \quad \dot{w}_n=0, \quad |n| \geq 2.
\]
Together with $v_n=-Vw_{+}+\sum_{k=-\infty}^n \dot{w}_k$ the latter yield
\beq
v_{-1}=v_0=-V w_{+}+\dfrac{w_{+}-w_c-\delta/2}{2(T-T_1)}, \quad v_n=-V w_{+}, \quad n \leq -2, \quad n \geq 1.
\label{eq:exact_CSW_vel}
\eeq

Using \eqref{eq:exact_CSW_strains} and \eqref{eq:exact_CSW_vel}, we obtain the renormalized energy \eqref{eq:Eren_D} given by
\[
E^D_{ren}(V)=\left(\dfrac{w_{+}-w_c-\delta/2}{2(T-T_1)}-V w_{+}\right)^2-V^2w_{+}^2+\Phi(w_0)+2\Phi(w_1)-3\Phi(w_+).
\]
The amplitude and renormalized energy of the obtained solution are shown by the dashed green curves in Fig.~\ref{fig:triCSWampen_DvsQC_alpha0}. One can see that this solution differs from the one for the QC model even at very small velocities, as illustrated in Fig.~\ref{fig:triCSW_DvsQC_alpha0}(b). Indeed, in the limit $V \to 0$ ($T \to \infty$), we have $w_0 \to w_c+\delta/2$, $w_n \to w_+$, $n \neq 0$, $v_n \to 0$ for all $n$,
and thus
\[
E^D_{ren}(0)=\Phi(w_0)-\Phi(w_+)=-(w_c-\delta/2+\beta\delta)(w_+-w_c-\delta/2).
\]
For comparison, we recall in the QC model the zero-velocity limit is given by \eqref{eq:sonic_limit}.
Thus the limiting renormalized energy \eqref{eq:Eren_QC} is given by
\[
E^{QC}_{ren}(0)=\int_{-\infty}^{\infty} \left(\Phi(w(x))-\Phi(w_+)\right)dx =-\dfrac{1}{\sqrt{3}}(w_c-\delta/2+\beta\delta)(w_+-w_c-\delta/2)=\dfrac{1}{\sqrt{3}}E^D_{ren}(0).
\]
In addition to the quantitative difference between discrete and QC solutions that persists to the sonic limit, it is important to note that for the discrete solution $w(\xi)-w_+$ is \emph{compact}, while its QC counterpart features exponential decay to the background strain.

\section{Stability of superkink solutions}
\label{sec:stab_super}
We tested stability of the superkinks by conducting numerical simulations of \eqref{eq:FPUstrain} on a finite chain. In the first set of simulations, we extracted initial conditions from the computed superkink solutions and used the corresponding fixed boundary conditions. These simulations resulted in steady propagation of the traveling wave with velocity that remained within $O(10^{-8})$ or less from the prescribed value for the entire range of velocities, suggesting that the traveling waves are at least long-lived and likely stable.

The second set of simulations was conducted on a chain with $L$ particles using free-end boundary conditions and Riemann initial data
\beq
w_n(0)=\begin{cases}w^{l}, & 1 \leq n \leq L/2\\
                    w^{r}, & L/2+1 \leq n \leq L
       \end{cases},
\qquad \dot{w}_n(0)=0, \quad n=1,\dots,L.
\label{eq:Riemann_data}
\eeq
The size $L$ of the chain was chosen sufficiently large to avoid any boundary effects.

Some results of simulations with Riemann data \eqref{eq:Riemann_data} are shown in Fig.~\ref{fig:triRiemann1} and Fig.~\ref{fig:triRiemann2},
where we fix $\beta=6$, $\delta=0.4$, $w_c=1$ and vary $\alpha$, $w^l$ and $w^r$.
\begin{figure}
\centering
\includegraphics[width=\textwidth]{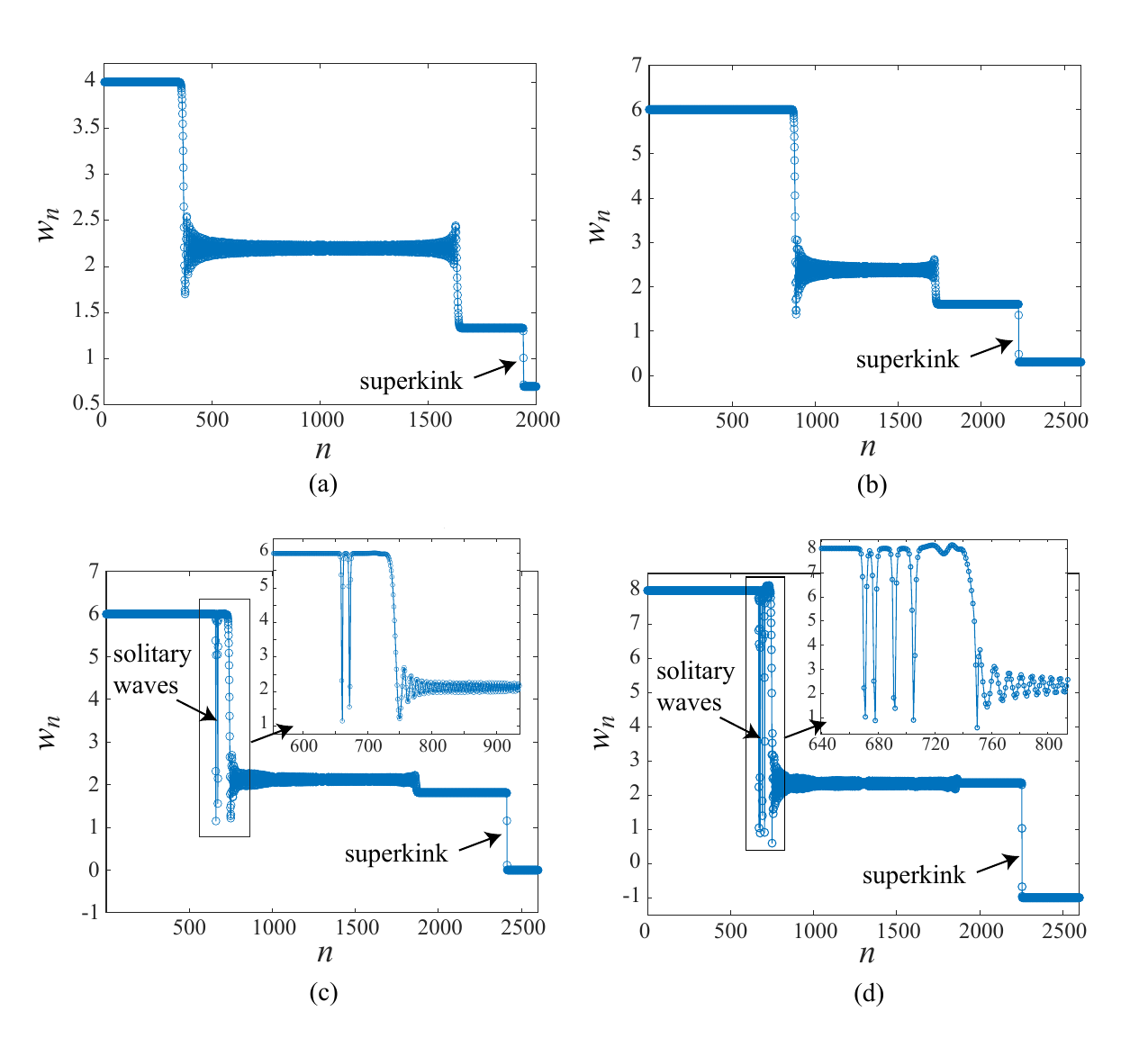}
\caption{\footnotesize Snapshots of strain profiles in simulations with Riemann initial data \eqref{eq:Riemann_data}: (a) $\alpha=2$, $w^l=4$, $w^r=0.7$, $L=2000$, $t=450$; (b) $\alpha=0.5$, $w^l=6$, $w^r=0.3$, $L=2600$, $t=600$; (c) $\alpha=0.5$, $w^l=6$, $w^r=0$, $L=2600$, $t=800$; (d) $\alpha=0.5$, $w^l=8$, $w^r=-1$, $L=2600$, $t=800$. Here $\beta=6$, $\delta=0.4$, $w_c=1$.}
\label{fig:triRiemann1}
\end{figure}
Typical scenario, where a superkink propagating to the right is trailed by linear dispersive shock waves moving in opposite directions with velocities $\pm \sqrt{\alpha}$, is shown in Fig.~\ref{fig:triRiemann1}(a) ($\alpha=2$, $w^l=4$, $w^r=0.7$) and Fig.~\ref{fig:triRiemann1}(b) ($\alpha=0.5$, $w^l=6$, $w^r=0.3$). The velocity of the superkink is $V_{SK}=2.0901$ in Fig.~\ref{fig:triRiemann1}(a) and $V_{SK}=1.5404$ in Fig.~\ref{fig:triRiemann1}(b), in agreement with \eqref{eq:Vsk1} when $w_+=w^r$, $V=V_{SK}$ for each case. A more complex dynamics is observed in Fig.~\ref{fig:triRiemann1}(c) ($\alpha=0.5$, $w^l=6$, $w^r=0$). In this case two solitary waves form behind the left dispersive shock wave and eventually move to the left with velocities $V_1=-0.8092$ and $V_2=-0.792$, while a superkink moves to the right (ahead of another dispersive shock wave) with velocity $V=1.39$, again consistent with \eqref{eq:Vsk1}. At $\alpha=0.5$, $w^l=8$, $w^r=-1$, the dynamics, shown in Fig.~\ref{fig:triRiemann1}(d), is similar but there are four solitary waves moving to the left, with velocities $V_1=-0.7913$, $V_2=-0.7891$, $V_3=-0.792$, $V_4=-0.7514$ near the end of the simulation, while a superkink propagates to the right with $V_{SK}=1.1926$.

The most interesting dynamics takes place in the case $\alpha=0$, as illustrated in Fig.~\ref{fig:triRiemann2}(a,b), where $w^l=6$ and $w^r=0.3$. In this case of inelastic red linear regime, we see a superkink propagating with $V_{SK}=1.5263$, in agreement with \eqref{eq:Vsk1}. Behind the superkink there is another transition front that moves to the left and connects the constant strain $w^l$ to a periodic train of solitary waves. The solitary wave train in this dispersive structure, known as a Whitham shock \cite{Sprenger20}, is spreading, with the left edge (the transition front) moving with velocity $V_L=-0.3415$, while its right edge propagates with $V_R=1.5022$, trailing the superkink. Similar dynamics is observed when we set $w^l=-0.75$, while keeping all other parameters the same; see Fig.~\ref{fig:triRiemann2}(c,d). In this case we have $V_L=-0.5003$, $V_R=1.1797$, $V_{SK}=1.2018$. The corresponding space-time evolution is shown in Fig.~\ref{fig:spacetime_alpha0}. While supporting stability of superkinks, these results also reveal the interesting phenomenon of Whitham shocks that, to our knowledge, have not been previously observed for the FPU system.
\begin{figure}
\centering
\includegraphics[width=\textwidth]{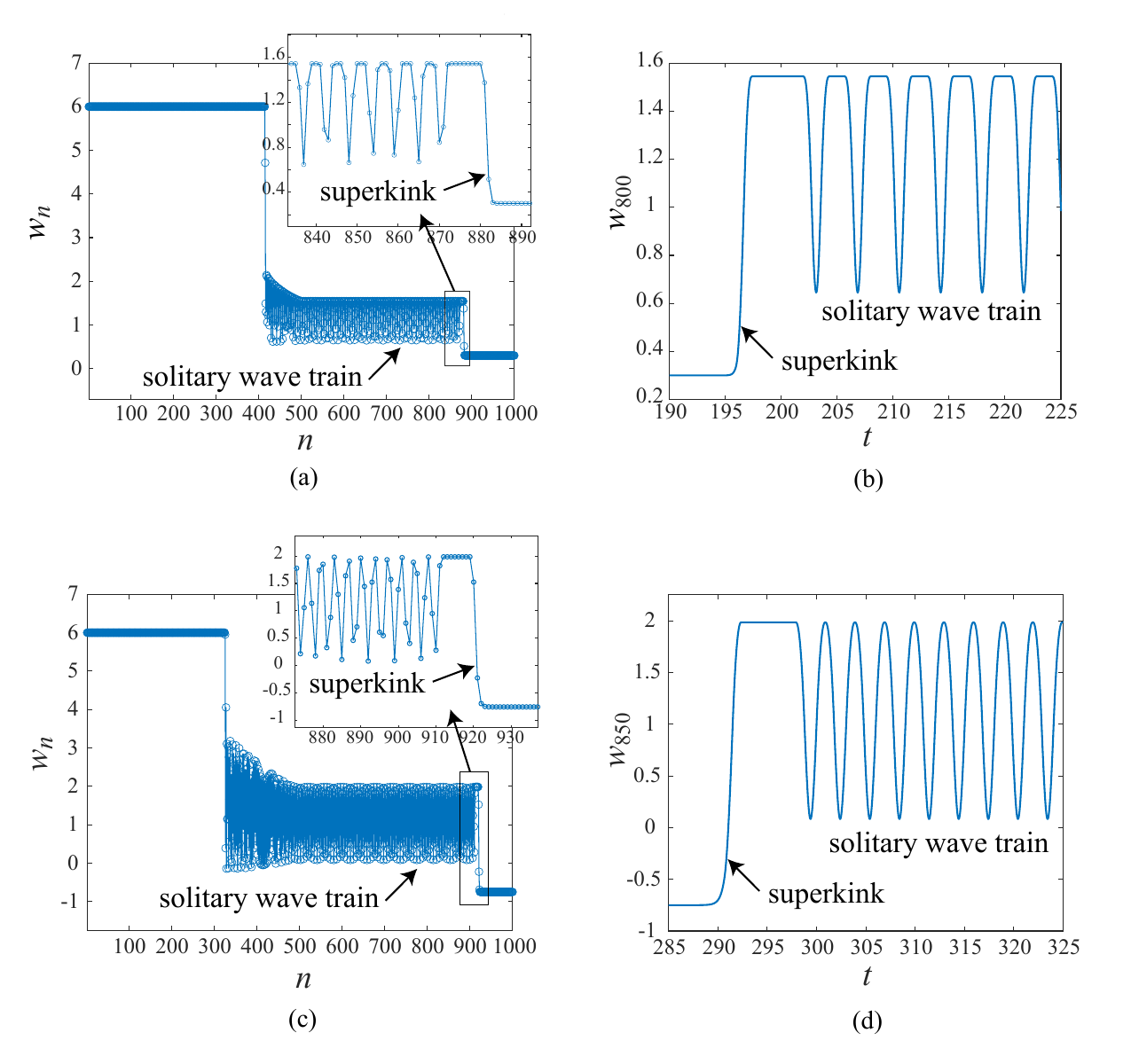}
\caption{\footnotesize (a) Snapshot of strain profiles in simulations with Riemann initial data \eqref{eq:Riemann_data} and $\alpha=0$, $w^l=6$, $w^r=0.3$, $L=1000$, $t=250$; (b) time evolution of $w_{800}(t)$ for the simulation in (a); (c) snapshot of strain profiles in simulations with Riemann initial data \eqref{eq:Riemann_data} and $\alpha=0$, $w^l=6$, $w^r=-0.75$, $L=1000$, $t=350$; (d) time evolution of $w_{850}(t)$ for the simulation in (c). Here $\beta=6$, $\delta=0.4$, $w_c=1$.}
\label{fig:triRiemann2}
\end{figure}
\begin{figure}
\centering
\includegraphics[width=\textwidth]{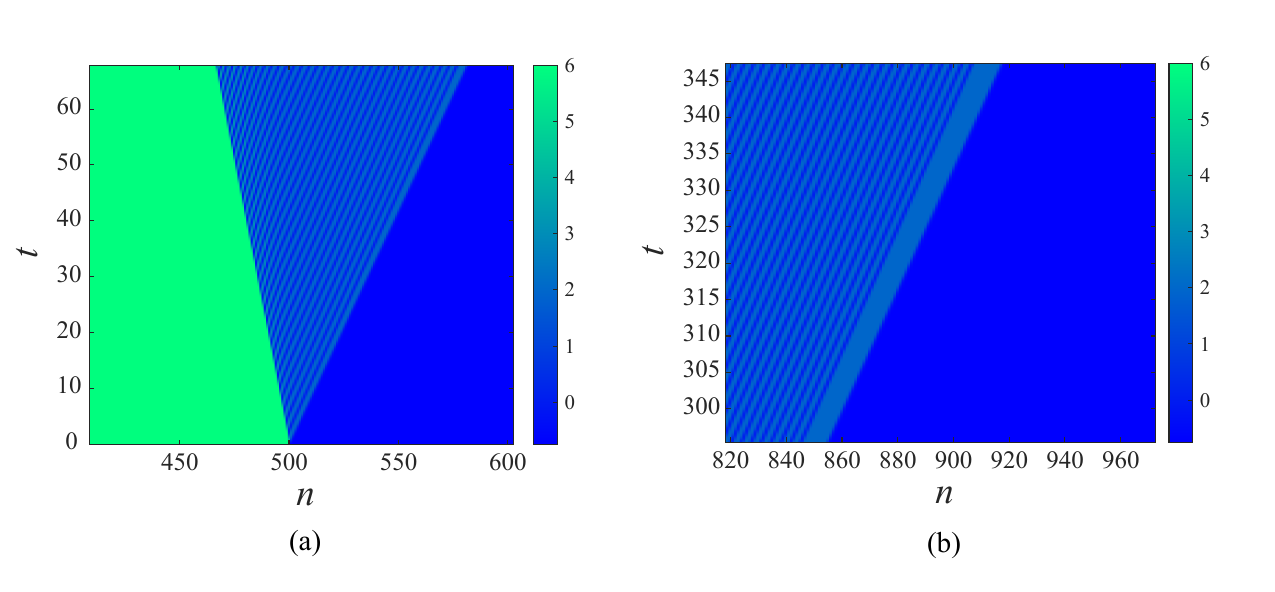}
\caption{\footnotesize Space-time evolution of strain in simulations with Riemann initial data \eqref{eq:Riemann_data}: (a) formation of the superkink front and SW train early in the simulation; (b) evolution near the end of the simulation, zoomed in around the superkink front.  Here $\alpha=0$, $\beta=6$, $\delta=0.4$, $w_c=1$, $w^l=6$, $w^r=-0.75$, $L=1000$.}
\label{fig:spacetime_alpha0}
\end{figure}

\section{Stability of solitary wave solutions}
\label{sec:stab_SW}
To investigate the linear stability of the obtained SW solutions in the case $\alpha>0$, we follow the approach in \cite{Cuevas17,Xu18,Vainchtein20} and use Floquet analysis that exploits periodicity-modulo-shift \eqref{eq:period} of the traveling wave solutions. Substituting $w_n(t)=\hat{w}_n(t)+\epsilon y_n(t)$ into \eqref{eq:FPUstrain}, where $\hat{w}_n(t)=w(n-Vt)$ is the traveling wave solution, and considering $O(\epsilon)$ terms, we obtain the governing equations for the linearized problem:
\beq
\ddot{y}_n=f'(\hat{w}_{n+1})y_{n+1}-2f'(\hat{w}_n)y_n+f'(\hat{w}_{n-1})y_{n-1}.
\label{eq:linear}
\eeq
The Floquet multipliers $\mu$ for this problem are the eigenvalues of the monodromy matrix $\mathcal{M}$ defined by
\begin{equation}
\label{eq:linear_map}
\left[\begin{array}{c}
  \{y_{n+1}(T)\} \\ \{\dot y_{n+1}(T)\} \\  \end{array}\right]
  =\mathcal{M}
  \left[\begin{array}{c}
  \{y_{n}(0)\} \\ \{\dot y_{n}(0)\} \\  \end{array}\right].
\end{equation}
To obtain $\mathcal{M}$, we compute the fundamental solution matrix $\mathbf{\Psi}(T)$, which maps $[\{y_{n}(0)\}, \{\dot y_{n}(0)\}]^T$ onto $[\{y_{n}(T)\}, \{\dot y_{n}(T)\}]^T$, $n=-N/2,\dots,N/2-1$, for the first-order linear system equivalent to \eqref{eq:linear}. We use periodic boundary conditions $y_{N/2}(t)=y_{-N/2}(t)$, $y_{-N/2-1}(t)=y_{N/2-1}(t)$, which is justified by the fact that for solitary waves the values $f'(\hat{w}_n)$ at the two ends of the chain have the same constant value (note, however, that this is not the case for superkinks unless $\alpha=1$). Due to the piecewise linear nature of \eqref{eq:trilinear}, the computation of the fundamental solution matrix $\Psi(T)$ involves
determining the times instances $T_i$, $i=1,\dots,k$, at which one of the nodes switches from one linear regime to another over the time interval $[0,T]$. This yields
\[
\Psi(T)=e^{\mathbf{C}_{k+1}(T-T_k)}e^{\mathbf{C}_k (T_k-T_{k-1})}\dots e^{\mathbf{C}_2(T_2-T_1)}e^{\mathbf{C}_1 T_1},
\]
where $\mathbf{C}_i$ has the block form
\[
\mathbf{C}_i=\left[\begin{array}{cc} \mathbf{O} & \mathbf{I}\\\mathbf{A}_i & \mathbf{O}\end{array}\right], \quad i=1,\dots,k+1
\]
involving the $N \times N$ identity matrix $\mathbf{I}$, the $N \times N$ zero matrix $\mathbf{O}$ and $N \times N$ matrices $\mathbf{A}_i$ that contain the coefficients for the corresponding linear system and have a tridiagonal structure extended to the upper right and lower left corner entries according to the periodic boundary conditions. We then shift the rows of $\mathbf{\Psi}(T)$ up by one row in the two parts of the matrix corresponding to $y_n$ and $\dot{y}_n$, respectively, with the last row in each part replaced by the first, obtaining $\mathcal{M}$ in \eqref{eq:linear_map}. We note that this procedure relies on the periodic boundary conditions, which, as mentioned above, are justified for solitary waves but in general not for superkinks.

The Floquet multipliers are related to the eigenvalues $\lambda$ of the linearization operator via $\mu=e^{\lambda/V}$, and thus $|\mu|>1$ ($\text{Re}(\lambda)>0$) corresponds to instability. The Hamiltonian nature of the problem means that there are quadruples of non-real Floquet multipliers, i.e., if $\mu$ is a multiplier, than so are $\bar{\mu}$, $1/\mu$ and $1/\bar{\mu}$, while the real multipliers come in pairs $\mu$ and $1/\mu$. Linear stability thus requires that all Floquet multipliers lie on the unit circle: $|\mu|=1$.

The resulting maximum modulus of Floquet multipliers as a function of $V$ is shown in Fig.~\ref{fig:floquet_triSW}.
\begin{figure}
\centering
\includegraphics[width=\textwidth]{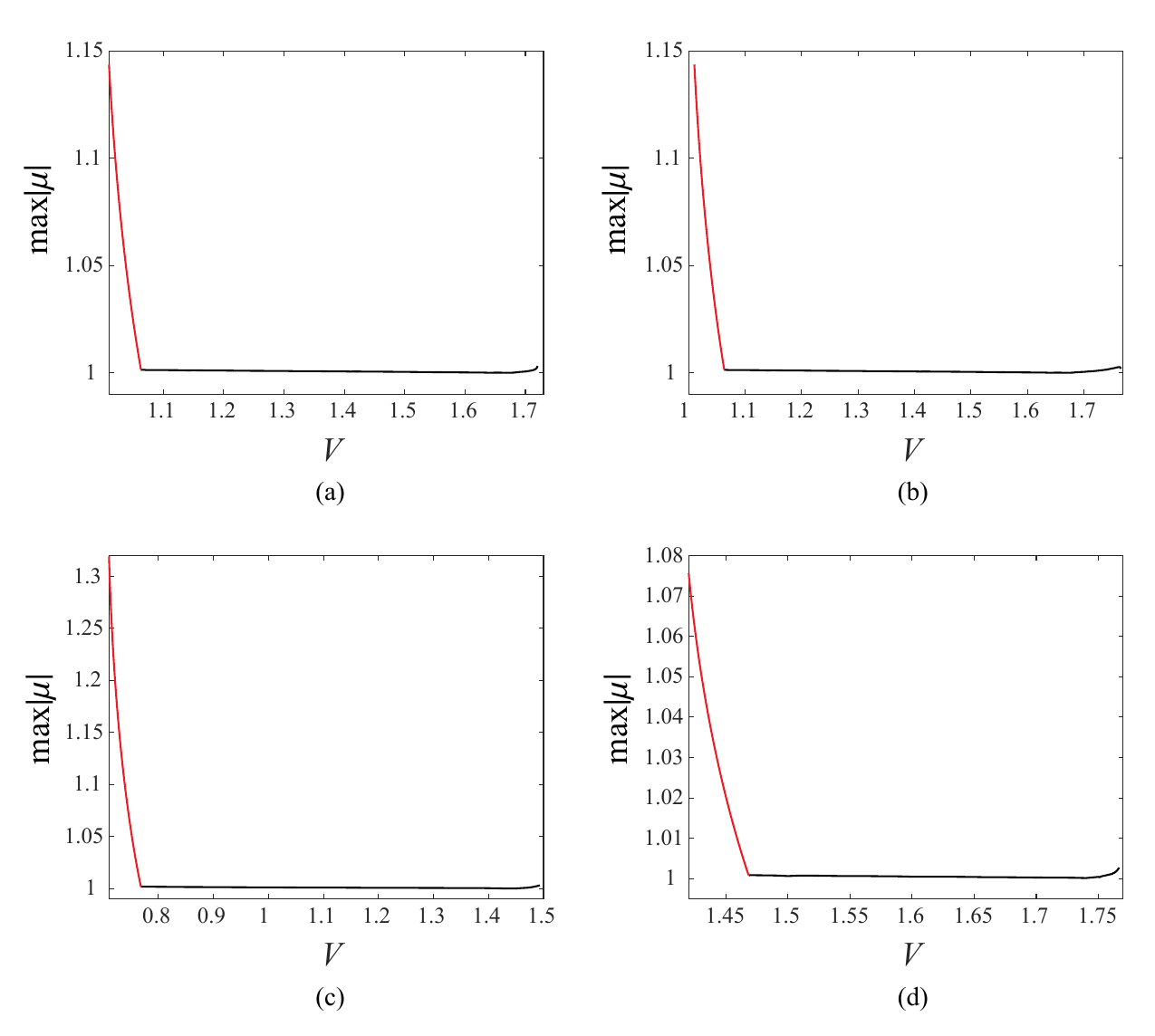}
\caption{\footnotesize Maximum modulus of Floquet multipliers for (a) tensile waves with $w_+=0.5$, $\alpha=0.5$; (b) tensile waves with $w_+=0.5$, $\alpha=2$; (c) compressive waves with $w_+=1.66$, $\alpha=0.5$; (d) compressive waves with $w_+=1.66$, $\alpha=2$. The red segments correspond to real multiplier $\mu>1$. Here $\beta=6$, $\delta=0.4$, $w_c=1$, and the corresponding values of the superkink velocity are $V_{SK}=1.72044$ in (a), $V_{SK}=1.76613$ in (b), $V_{SK}=1.49541$ in (c) and $V_{SK}=1.76697$ in (d).}
\label{fig:floquet_triSW}
\end{figure}
\begin{figure}
\centering
\includegraphics[width=\textwidth]{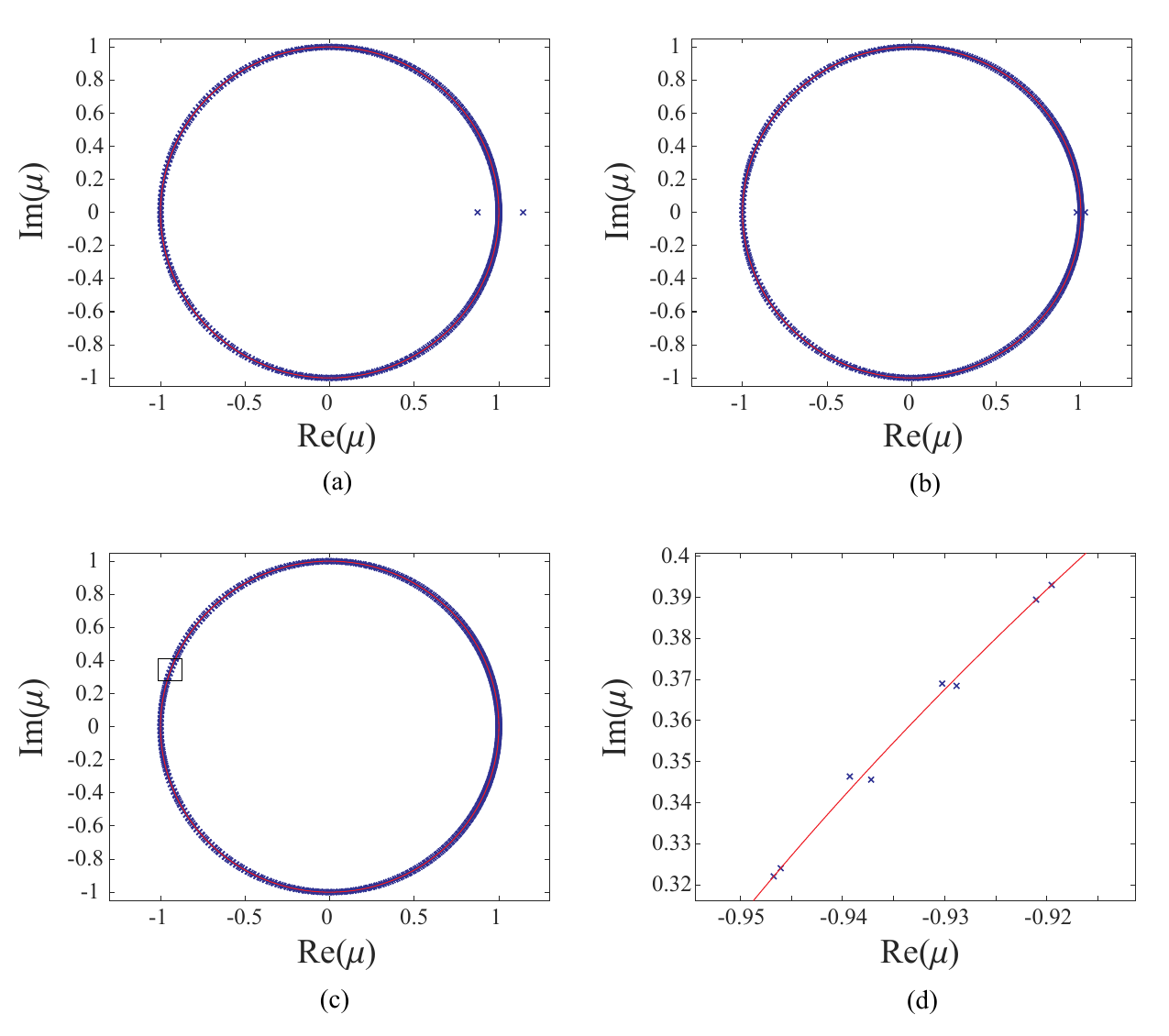}
\caption{\footnotesize Floquet multipliers (blue crosses) for tensile solitary waves at (a) $V=1.01$; (b) $V=1.05$; (c) $V=1.14$; (d) $V=1.14$, zooming inside the rectangle in panel (c). The unit circle is marked in red. Here $\alpha=2$, $w_+=0.5$, $\beta=6$, $\delta=0.4$, $w_c=1$.}
\label{fig:floquet_examples}
\end{figure}
\begin{figure}
\centering
\includegraphics[width=\textwidth]{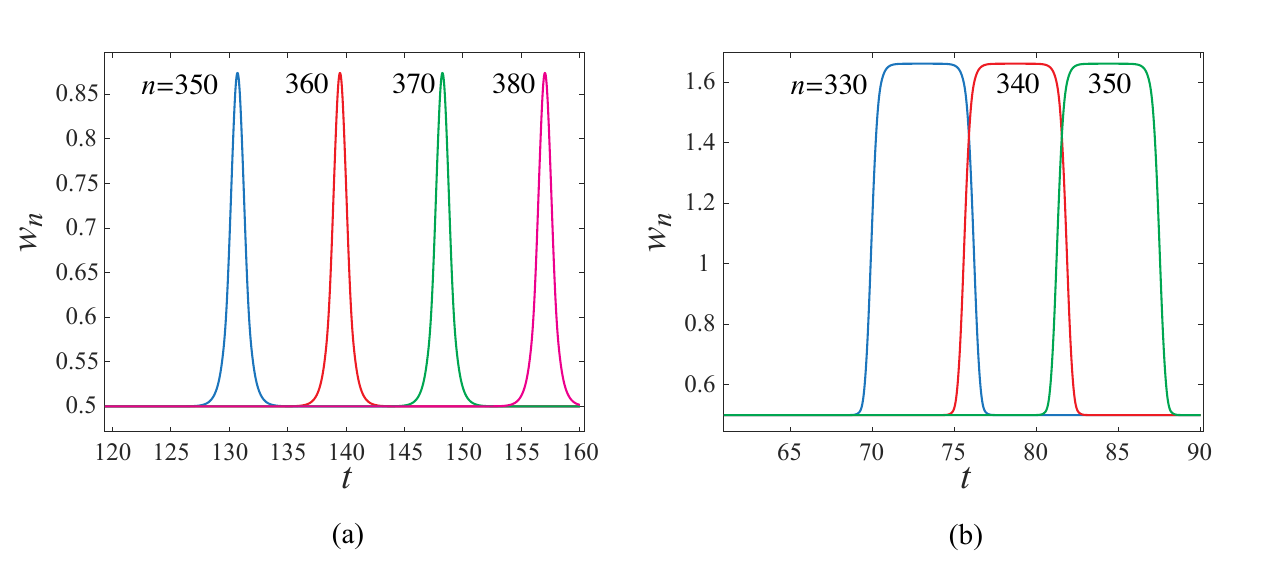}
\caption{\footnotesize Strain evolution in numerical simulations initiated by computed solitary waves with (a) $V=1.14$; (b) $V=1.7661272$. Here $\alpha=2$, $w_+=0.5$, $\beta=6$, , $\delta=0.4$, $w_c=1$.}
\label{fig:triTSWstab_examples}
\end{figure}
In each case, for velocities below a certain threshold $V_*$ the maximum-modulus multiplier is real $\mu>1$ (red segments in Fig.~\ref{fig:floquet_triSW}) and corresponds to exponential instability mode. As the threshold velocity $V_*$ is approached, the real multiplier $\mu$ outside the unit circle and the companion real multiplier $1/\mu$ inside the unit circle move toward the unit circle and join it at $V=V_*$. This is illustrated in Fig.~\ref{fig:floquet_examples} for the case $\alpha=2$, $w_+=0.5$, where we see the two real multipliers approach the unit circle as velocity is increased from $1.01$ (panel (a)) to $1.05$ (panel (b)), slightly below $V_* \approx 1.063$. Above $V_*$ the maximum-modulus multipliers are complex and correspond to mild instability modes similar to those observed in \cite{Marin98,VT24,Xu18}. Since their magnitude decreases as the chain size is increased, these mild instabilities appear to be a spurious artifact of the finite chain size. An example is shown in Fig.~\ref{fig:floquet_examples}(c,d). Direct numerical simulations initiated by solitary waves with velocities $V>V_*$ show steady propagation of the waves and suggest their effective stability, as illustrated in Fig.~\ref{fig:triTSWstab_examples}.

For solitary waves that tend to zero at infinity, the onset of exponential instability associated with $\mu>1$ typically corresponds to threshold velocities at which their energy changes monotonicity \cite{Cuevas17,Xu18}. In this case, however, the energy of the wave is infinite, and the onset of instability occurs prior to the change in monotonicity of the renormalized energy \eqref{eq:Eren_D}, which takes place at $V>V_*$, as can be seen in Fig.~\ref{fig:triTSWenergy_DvsQC} and Fig.~\ref{fig:triCSWenergy_DvsQC}(b,c). It is possible that another relevant quantity changes monotonicity at $V=V_*$.

To explore the consequences of the instability at $V<V_*$, we ran numerical simulations initiated by an unstable solitary waves perturbed along the corresponding eigenmode. A typical scenario for ensuing dynamic evolution is shown in Fig.~\ref{fig:triCSWpert}, where the simulation was initiated by the perturbed unstable compressive wave with velocity $V=0.72$ below the threshold value
$V_* \approx 0.7685$ at $\alpha=0.5$, $w_{+}=1.66$ (see Fig.~\ref{fig:floquet_triSW}(c) for the corresponding maximum-modulus Floquet multipliers). One can see formation of an apparently stable solitary wave with $V=0.8288$ above the threshold followed by dispersive wave that propagates with lower (sonic) speed.
\begin{figure}
\centering
\includegraphics[width=\textwidth]{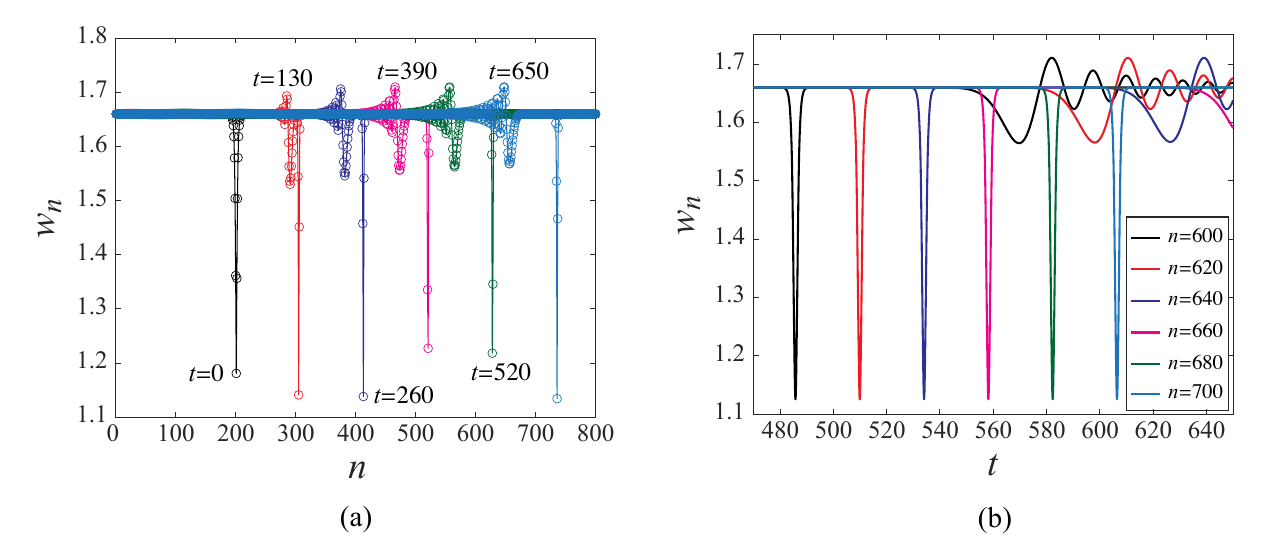}
\caption{\footnotesize (a) Snapshots of strain profiles and (b) $w_n(t)$ at different fixed $n$ for the dynamic evolution initiated by an unstable compressive wave propagating with velocity $V=0.72<V_*$ at $\alpha=0.5$ and $w_{+}=1.66$, perturbed along the eigenmode corresponding to the real Floquet multiplier $\mu=1.20795$. The dynamic evolution leads to the formation of compressive solitary wave with $V=0.8288>V_*$ followed by a sonic dispersive wave. Here $V_* \approx 0.7685$, $\delta=0.4$, $\beta=6$, $w_c=1$.}
\label{fig:triCSWpert}
\end{figure}

Stability of compressive waves in the degenerate case $\alpha=0$, shown in Fig.~\ref{fig:triCSW_DvsQC}, was investigated numerically.
Simulation results suggest stable propagation of solitary waves in the entire velocity range $0<V<V_{SK}$.

\section{Concluding remarks}
\label{sec:conclusions}

In  this paper, we considered the FPU system with trilinear force-elongation relation. It was chosen to be of generally asymmetric soft-hard-soft type, and the resulting mechanical behavior can be classified as hardening-softening. We showed that in addition to the classical finite-amplitude, spatially  localized solitary waves, this model exhibits supersonic kinks (superkinks) and finite-amplitude, spatially delocalized flat-top solitary waves which acquire the structure of a kink-antikink bundle when their velocity tends to the kink limit. Exploiting the periodic-modulo-shift property of traveling waves, we computed these solutions as fixed points of the corresponding nonlinear map. In a particularly interesting degenerate case when the elastic modulus of one of the soft regimes is zero, we obtained exact solutions for sufficiently slow solitary waves.

Floquet analysis of solitary waves in the non-degenerate case shows that near-sonic waves are unstable when their velocity is below a certain threshold. Perturbation along the corresponding eigenmode led to the formation of a stable wave with velocity above the threshold. Stability of these and other solutions was also confirmed by direct numerical simulations initiated by the computed traveling waves and piecewise constant Riemann data. In the degenerate case, Riemann simulations revealed emergence of Whitham shocks involving
periodic train of solitary waves.

We also analyzed in detail a QC approximation of the FPU problem, which introduces into the continuum model mixed space-time higher-order derivative term describing microinertia. Using this model, we derived explicit solutions for both solitary waves and superkinks.  The analytical transparency of the QC model allowed us to examine in full detail the properties of the waves and the effect of asymmetry of the interaction force. Comparison of the obtained solutions with their discrete counterparts showed that the QC model captures the main
effects qualitatively, and quantitative agreement exists near the superkink and, for the non-degenerate case, sonic limit. In the degenerate case the discrepancy between the discrete and QC model persists to the sonic limit of zero velocity.

Finally, we mention a potential application of the obtained results. It is known that persistent ``particle-like'' wave packets can be generated in mechanical metamaterials  to transfer mechanical energy and communicate mechanical information \cite{bertoldi2017flexible}. An important class of applications of such metamaterials involves autonomous locomotion. In particular, wave-driven robots, utilizing  geometric phase transitions as internal mechanisms, are becoming a subject of intense research due to their adaptability to complex environments \cite{deng2020pulse}. Moreover, a new paradigm in robotics is emerging in the form of a transition from movable machines to movable materials with self-propulsion interpreted as dynamics of uniform regions (or domains) bounded by coherently moving interfaces (domain walls) \cite{yasuda2020transition}. The idea is that constructive interplay  between material nonlinearity and dispersion  can  lead to the emergence of such robust disturbances which would propagate with constant velocity and fixed profile.  The  corresponding nonlinear wave  would be then  associated with some functionality, as, for instance, in the case of peristaltic motion \cite{Gorbushin21}. The advantage of soft mechanical alternatives to rigid controlling  actuators in otherwise soft crawling robots is obvious, and the main challenge is to learn to generate such programmable  dynamic regimes. The delocalized flat-top solitary waves discussed in this work offer an example of stable nonlinear pulses  which can be used to make the corresponding  metamaterial crawl. In this respect our conclusion that the resulting delocalized active pulses are necessarily supersonic is still realistic because in biologically relevant soft solids the acoustic speeds may be arbitrarily small. Note also that in this perspective our trilinear model  can be interpreted as describing a material capable of generating active stresses \cite{Gorbushin19}.  In the same spirit, the flat-top solitary waves would imply dynamic passive-to-active transformations taking place in the front of a steadily moving pulse with the corresponding reverse transformation taking place in its rear.

\begin{section}*{Acknowledgements}
 The work of AV was supported by the NSF grant DMS-2204880. LT  acknowledges the support of the French Agence Nationale de la Recherche under the grant ANR-17-CE08-0047-02.
\end{section}

\appendix
\numberwithin{equation}{section}
 \setcounter{equation}{0}

\section{Some technical results}
\label{app:coefs}
The coefficients in \eqref{eq:TSW2} are found by imposing continuity of $w(\xi)$ and $w'(\xi)$. This yields
\beq
\begin{split}
A&=(q(q(w_S-w_{+}) \cosh(s z_2) \sin(q(z_1 - z_2)) + s (w_S-w_{-} \\
&+ (w_{+} - w_S) \cos(q(z_1 - z_2))) \sinh(s z_2)))/\mathcal{D},\\
B&=(q r (w_S-w_{+}) \cos(q z_2) \cosh(s z_2) + s ((w_S-w_{-}) (q \cos(q z_1) + r \sin(q z_1)) \\
&-r (w_S-w_{+}) \sin(q z_2)) \sinh(s z_2))/\mathcal{D},\\
C&=(q r (w_S-w_+) \cosh(s z_2) \sin(q z_2) + s (r (w_S-w_+) \cos(q z_2) \\
&- (w_S-w_{-})(r \cos(q z_1) - q \sin(q z_1)))\sinh(s z_2))/\mathcal{D},\\
D&=(q (r (w_S-w_+) + (w_{-} - w_S) (r \cos[q(z_1 - z_2)) - q \sin(q (z_1 - z_2)))))/\mathcal{D},
\end{split}
\label{eq:TSW2_coefs}
\eeq
where
\[
\mathcal{D}=\sin(q (z_1 - z_2)) (q^2 \cosh(s z_2) - r s \sinh(s z_2))-q \cos(q(z_1 - z_2))(r\cosh(s z_2) + s \sinh(s z_2)).
\]
Similarly, we find the coefficients in \eqref{eq:CSW2}:
\beq
\begin{split}
A&=q(q(w_S-w_{+})\cosh(r z2) \sin(q(z_1 - z_2)) + r (w_S-w_{-}\\
&+(w_+ - w_S) \cos(q (z_1 - z_2))) \sinh(r z_2)))/\mathcal{D}\\
B&=(q s (w_S-w_{+})\cos(q z_2) \cosh(r z_2) + r (w_S-w_{-})(q \cos(q z_1) + s \sin(q z_1))\\
&+s(w_{+} - w_S)\sin(q z_2))\sinh(r z_2))/\mathcal{D}\\
C&=(q s(w_S-w_{+})\cosh(r z_2) \sin(q z_2) + r(s (w_S-w_+)\cos(q z_2)\\
&+(w_{-} - w_S) (s \cos(q z_1) - q \sin(q z_1)))\sinh(r z_2))/\mathcal{D}\\
D&=(q(s(w_S-w_{+}) + (w_{-} - w_S) (s\cos(q(z_1 - z2)) -q\sin(q(z_1 - z_2)))))/\mathcal{D},
\end{split}
\label{eq:CSW2_coefs}
\eeq
where
\[
\mathcal{D}=\sin(q(z_1- z_2))(q^2\cosh(r z_2) - r s \sinh(r z_2))-q \cos(q (z_1 - z_2)) (s\cosh(r z_2) + r\sinh(r z_2)).
\]


\begin{thebibliography}{10}

\bibitem{yasuda2020transition}
H.~Yasuda, L.~M. Korpas, and J.~R. Raney.
\newblock Transition waves and formation of domain walls in multistable
  mechanical metamaterials.
\newblock {\em Phys. Rev. Appl.}, 13(5):054067, 2020.

\bibitem{raney2016stable}
J.~R. Raney, N.~Nadkarni, C.~Daraio, D.~M. Kochmann, J.~A. Lewis, and
  K.~Bertoldi.
\newblock Stable propagation of mechanical signals in soft media using stored
  elastic energy.
\newblock {\em Proc. Nat. Acad. Sci.}, 113(35):9722--9727, 2016.

\bibitem{kochmann2017exploiting}
D.~M. Kochmann and K.~Bertoldi.
\newblock Exploiting microstructural instabilities in solids and structures:
  from metamaterials to structural transitions.
\newblock {\em Appl. Mech. Rev.}, 69(5), 2017.

\bibitem{zhang2019programmable}
Y.~Zhang, B.~Li, Q.~S. Zheng, G.~M. Genin, and C.~Q. Chen.
\newblock Programmable and robust static topological solitons in mechanical
  metamaterials.
\newblock {\em Nat. Comm.}, 10(1):1--8, 2019.

\bibitem{bertoldi2017flexible}
K.~Bertoldi, V.~Vitelli, J.~Christensen, and M.~{Van Hecke}.
\newblock Flexible mechanical metamaterials.
\newblock {\em Nature Rev. Mater.}, 2(11):1--11, 2017.

\bibitem{yasuda2019origami}
H.~Yasuda, Y.~Miyazawa, E.~G. Charalampidis, C.~Chong, P.~G. Kevrekidis, and
  J.~Yang.
\newblock Origami-based impact mitigation via rarefaction solitary wave
  creation.
\newblock {\em Sci. Adv.}, 5(5):eaau2835, 2019.

\bibitem{ZK65}
N.~J. Zabusky and M.~D. Kruskal.
\newblock Interaction of ``solitons" in a collisionless plasma and the
  recurrence of initial states.
\newblock {\em Phys. Rev. Lett.}, 15(6):240--243, 1965.

\bibitem{fermi1955studies}
E.~Fermi, P.~Pasta, S.~Ulam, and M.~Tsingou.
\newblock Studies of the nonlinear problems.
\newblock Technical report, Los Alamos National Laboratory, Los Alamos, NM,
  USA, 1955.

\bibitem{berman2005fermi}
G.~P. Berman and F.~M. Izrailev.
\newblock The {Fermi--Pasta--Ulam} problem: fifty years of progress.
\newblock {\em Chaos}, 15(1):015104, 2005.

\bibitem{gallavotti2007fermi}
G.~Gallavotti.
\newblock {\em The {Fermi--Pasta--Ulam} problem: a status report}, volume 728.
\newblock Springer, 2007.

\bibitem{remoissenet2013waves}
M.~Remoissenet.
\newblock {\em Waves called solitons: concepts and experiments}.
\newblock Springer Science \& Business Media, 2013.

\bibitem{newell1985solitons}
A.~C. Newell.
\newblock {\em Solitons in mathematics and physics}.
\newblock SIAM, 1985.

\bibitem{fokas2012important}
A.~S. Fokas and V.~E. Zakharov.
\newblock {\em Important developments in soliton theory}.
\newblock Springer Science \& Business Media, 2012.

\bibitem{Vainchtein22}
A.~Vainchtein.
\newblock Solitary waves in {FPU}-type lattices.
\newblock {\em Physica D}, page 133252, 2022.

\bibitem{ablowitz2011nonlinear}
M.~J. Ablowitz.
\newblock {\em Nonlinear dispersive waves: asymptotic analysis and solitons},
  volume~47.
\newblock Cambridge University Press, 2011.

\bibitem{Yasenchuk21}
Y.~F Yasenchuk, E.~S. Marchenko, S.~V. Gunter, G.~A. Baigonakova, O.~V.
  Kokorev, A.~A. Volinsky, and E.~B. Topolnitsky.
\newblock Softening effects in biological tissues and {NiTi} knitwear during
  cyclic loading.
\newblock {\em Materials}, 14(21):6256, 2021.

\bibitem{Sensini18}
A.~Sensini and L.~Cristofolini.
\newblock Biofabrication of electrospun scaffolds for the regeneration of
  tendons and ligaments.
\newblock {\em Materials}, 11(10):1963, 2018.

\bibitem{Millereau18}
P.~Millereau, E.~Ducrot, J.~M. Clough, M.~E. Wiseman, H.~R. Brown, R.~P.
  Sijbesma, and C.~Creton.
\newblock Mechanics of elastomeric molecular composites.
\newblock {\em Proc. Nat. Acad. Sci.}, 115(37):9110--9115, 2018.

\bibitem{Iooss00}
G.~Iooss.
\newblock Travelling waves in the {Fermi-Pasta-Ulam} lattice.
\newblock {\em Nonlinearity}, 13(3):849, 2000.

\bibitem{HerrmannRademacher10}
M.~Herrmann and J.~D.~M. Rademacher.
\newblock Heteroclinic travelling waves in convex {FPU}-type chains.
\newblock {\em SIAM J. Math. Anal.}, 42(4):1483--1504, 2010.

\bibitem{Herrmann11}
M.~Herrmann.
\newblock Action minimising fronts in general {FPU-type} chains.
\newblock {\em J. Nonlin. Sci.}, 21(1):33--55, 2011.

\bibitem{Gorbushin19}
N.~Gorbushin and L.~Truskinovsky.
\newblock Supersonic kinks and solitons in active solids.
\newblock {\em Phil. Trans. Royal Soc. A}, 378(2162):20190115, 2019.

\bibitem{Gorbushin21}
N.~Gorbushin and L.~Truskinovsky.
\newblock Peristalsis by pulses of activity.
\newblock {\em Phys. Rev. E}, 103(4):042411, 2021.

\bibitem{VT24}
A.~Vainchtein and L.~Truskinovsky.
\newblock When discrete fronts and pulses form a single family: {FPU} chain
  with hardening-softening springs.
\newblock {\em Physica D}, 2024.
\newblock to appear.

\bibitem{Gorbushin22}
N.~Gorbushin, A.~Vainchtein, and L.~Truskinovsky.
\newblock Transition fronts and their universality classes.
\newblock {\em Phys. Rev. E}, 106(2):024210, 2022.

\bibitem{collins1981quasicontinuum}
M.~A. Collins.
\newblock A quasicontinuum approximation for solitons in an atomic chain.
\newblock {\em Chem. Phys. Lett.}, 77(2):342--347, 1981.

\bibitem{Rosenau86}
P.~Rosenau.
\newblock Dynamics of nonlinear mass-spring chains near the continuum limit.
\newblock {\em Phys. Let. A}, 118(5):222--227, 1986.

\bibitem{kevrekidis2002continuum}
P.~G. Kevrekidis, I.~G. Kevrekidis, A.~R. Bishop, and E.~S. Titi.
\newblock Continuum approach to discreteness.
\newblock {\em Phys. Rev. E}, 65(4):046613, 2002.

\bibitem{feng2004quasi}
B.-F. Feng, Y.~Doi, and T.~Kawahara.
\newblock Quasi-continuum approximation for discrete breathers in
  {Fermi--Pasta--Ulam} atomic chains.
\newblock {\em J. Phys. Soc. Japan}, 73(8):2100--2111, 2004.

\bibitem{Aubry09}
S.~Aubry and L.~Proville.
\newblock Pressure fronts in {1D} damped nonlinear lattices, 2009.
\newblock arXiv preprint arXiv:0910.4890.

\bibitem{Vainchtein20}
A.~Vainchtein, J.~Cuevas-Maraver, P.~G. Kevrekidis, and H.~Xu.
\newblock Stability of traveling waves in a driven {Frenkel--Kontorova} model.
\newblock {\em Commun. Nonlin. Sci. Numer. Simul.}, 85:105236, 2020.

\bibitem{James21}
G.~James.
\newblock Traveling fronts in dissipative granular chains and nonlinear
  lattices.
\newblock {\em Nonlinearity}, 34(3):1758, 2021.

\bibitem{Cuevas17}
J.~Cuevas-Maraver, P.~G. Kevrekidis, A.~Vainchtein, and H.~Xu.
\newblock Unifying perspective: solitary traveling waves as discrete breathers
  in {Hamiltonian} lattices and energy criteria for their stability.
\newblock {\em Phys. Rev. E}, 96(3):032214, 2017.

\bibitem{Xu18}
H.~Xu, J.~Cuevas-Maraver, P.~G. Kevrekidis, and A.~Vainchtein.
\newblock An energy-based stability criterion for solitary travelling waves in
  {Hamiltonian} lattices.
\newblock {\em Phil. Trans. Royal Soc. A}, 376(2117):20170192, 2018.

\bibitem{Serre07}
D.~Serre.
\newblock Discrete shock profiles: Existence and stability.
\newblock In A.~Bressan, D.~Serre, M.~Williams, K.~Zumbrun, and D.~Serre,
  editors, {\em Hyperbolic Systems of Balance Laws: {Lectures given at the CIME
  Summer School held in Cetraro, Italy, July 14--21, 2003}}, pages 79--158.
  Springer, 2007.

\bibitem{Trusk87}
L.~M. Truskinovskii.
\newblock Dynamics of non-equilibrium phase boundaries in a heat conducting
  non-linearly elastic medium.
\newblock {\em J. Appl. Math. Mech.}, 51(6):777--784, 1987.

\bibitem{Friesecke94}
G.~Friesecke and J.~A.~D. Wattis.
\newblock Existence theorem for solitary waves on lattices.
\newblock {\em Commun. Mat. Phys.}, 161(2):391--418, 1994.

\bibitem{RosenauOron20}
P.~Rosenau and A.~Oron.
\newblock Flatons: flat-top solitons in extended {Gardner-like} equations.
\newblock {\em Commun. Nonlin. Sci. Numer. Simul.}, 91:105442, 2020.

\bibitem{RosenauOron22}
P.~Rosenau and A.~Oron.
\newblock Compact patterns in a class of sublinear {Gardner} equations.
\newblock {\em Commun. Nonlin. Sci. Numer. Simul.}, 110:106384, 2022.

\bibitem{RosenauPikovsky20}
P.~Rosenau and A.~Pikovsky.
\newblock Solitary phase waves in a chain of autonomous oscillators.
\newblock {\em Chaos}, 30(5), 2020.

\bibitem{RosenauPikovsky21}
P.~Rosenau and A.~Pikovsky.
\newblock Waves in strongly nonlinear {Gardner-like} equations on a lattice.
\newblock {\em Nonlinearity}, 34(8):5872, 2021.

\bibitem{slepyan2001feeding}
L.~I. Slepyan.
\newblock Feeding and dissipative waves in fracture and phase transition: {II.
  Phase-transition} waves.
\newblock {\em J. Mech. Phys. Solids}, 49(3):513--550, 2001.

\bibitem{gorbushin2020frictionless}
N.~Gorbushin, G.~Mishuris, and L.~Truskinovsky.
\newblock Frictionless motion of lattice defects.
\newblock {\em Phys. Rev. Lett.}, 125(19):195502, 2020.

\bibitem{Nesterenko01}
V.~F. Nesterenko.
\newblock {\em Dynamics of Heterogeneous Materials}.
\newblock Springer, 2001.

\bibitem{Sen08}
S.~Sen, J.~Hong, J.~Bang, E.~Avalos, and R.~Doney.
\newblock Solitary waves in the granular chain.
\newblock {\em Phys. Rep.}, 462(2):21--66, 2008.

\bibitem{Chong17}
C.~Chong, M.~A. Porter, P.~G. Kevrekidis, and C.~Daraio.
\newblock Nonlinear coherent structures in granular crystals.
\newblock {\em J. Phys.}, 29(41):413003, 2017.

\bibitem{MarinAubry96}
J.~L. Mar\'{\i}n and S.~Aubry.
\newblock Breathers in nonlinear lattices: numerical calculation from the
  anticontinuous limit.
\newblock {\em Nonlinearity}, 9:1501--1528, 1996.

\bibitem{Sprenger20}
P.~Sprenger and M.~A. Hoefer.
\newblock Discontinuous shock solutions of the {Whitham} modulation equations
  as zero dispersion limits of traveling waves.
\newblock {\em Nonlinearity}, 33(7):3268, 2020.

\bibitem{Marin98}
J.~L. Mar{\'\i}n and S.~Aubry.
\newblock Finite size effects on instabilities of discrete breathers.
\newblock {\em Physica D}, 119(1-2):163--174, 1998.

\bibitem{deng2020pulse}
B.~Deng, L.~Chen, D.~Wei, V.~Tournat, and K.~Bertoldi.
\newblock Pulse-driven robot: {Motion} via solitary waves.
\newblock {\em Sci. Adv.}, 6(18):eaaz1166, 2020.

\end{thebibliography}

\end{document}